\documentclass{aa}

\usepackage[varg]{txfonts}
\usepackage{graphicx}
\usepackage{amsmath}
\usepackage{amssymb}
\usepackage{microtype}
\usepackage{booktabs}
\usepackage{multirow}
\usepackage{subcaption}
\usepackage{tablefootnote}
\usepackage{threeparttable}
\usepackage{hyperref}

\makeatletter
\renewcommand*\aa@pageof{, page \thepage{} of \pageref*{LastPage}}
\makeatother

\hypersetup{colorlinks=true, linkcolor=blue, citecolor=blue}

\begin{document}

    

    \title{Grain growth and its chemical impact in the first hydrostatic core phase}

    
    \author{D. Navarro-Almaida
          \inst{1}\fnmsep\thanks{\email{david.navarroalmaida@cea.fr}}
          \and
          U. Lebreuilly\inst{1}
          \and
          P. Hennebelle\inst{1}
          \and
          A. Fuente\inst{2}
          \and
          B. Commerçon\inst{3}
          \and
          R. Le Gal\inst{4}
          \and
          V. Wakelam\inst{5}
          \and
          M. Gerin\inst{6}
          \and
          P. Rivi\'ere-Marichalar\inst{7}
          \and
          L. Beitia-Antero\inst{8}
          \and
          Y. Ascasibar\inst{9}
          }

    \institute{
        Universit\'{e} Paris-Saclay, Universit\'e Paris Cit\'{e}, CEA, CNRS, AIM, F-91191 Gif-sur-Yvette, France
        \and
        Centro de Astrobiolog\'ia (CSIC/INTA), Ctra. de Torrej\'on a Ajalvir km 4, 28806 Torrej\'on de Ardoz, Spain
        \and
        Universit\'e de Lyon, ENS de Lyon, Universit\'e Lyon 1, CNRS, Centre de Recherche Astrophysique de Lyon UMR5574, 69007 Lyon, France
        \and
        Institut de Plan\'etologie et d'Astrophysique de Grenoble (IPAG), Universit\'e Grenoble Alpes, CNRS, F-38000 Grenoble, France
        \and
        Laboratoire d'Astrophysique de Bordeaux, Univ. Bordeaux, CNRS, B18N, all\'ee Geoffroy Saint-Hilaire, 33615 Pessac, France
        \and
        LERMA, Observatoire de Paris, PSL Research University, CNRS, Sorbonne Universit\'e 61, Avenue de l'Observatoire, 75014 Paris, France
        \and
        Observatorio Astron\'omico Nacional (OAN, IGN), Calle Alfonso XII, 3, 28014 Madrid, Spain
        \and
        Departamento de Estad\'istica e Investigaci\'on Operativa, Facultad de Ciencias Matem\'aticas, Universidad Complutense de Madrid, Plaza de las Ciencias, 3, 28040 Madrid, Spain
        \and
        Departamento de F\'isica Te\'orica, Universidad Aut\'onoma de Madrid, 28049 Madrid, Spain
        }
    \date{}
    
    \abstract{The first hydrostatic core (FHSC) phase is a brief stage in the protostellar evolution that is difficult to detect. Its chemical composition determine that of later evolutionary stages. Numerical simulations are the tool of choice to study these objects.}
    {Our goal is to characterize the chemical evolution of gas and dust during the formation of the FHSC. Moreover, we are interested in analyzing, for the first time with 3D magnetohydrodynamic (MHD) simulations, the role of grain growth in its chemistry.}
    {We postprocessed $2\times10^{5}$ tracer particles from a \texttt{RAMSES} non-ideal MHD simulation using the codes \texttt{NAUTILUS} and \texttt{SHARK} to follow the chemistry and grain growth throughout the simulation.}
    {Gas-phase abundances of most of the C, O, N, and S reservoirs in the hot corino at the end of the simulation match the ice-phase abundances from the prestellar phase. Interstellar complex organic molecules (iCOMs) such as methyl formate, acetaldehyde, and formamide are formed during the warm-up process. Grain size in the hot corino $(n_{\rm H}>10^{11}\ {\rm cm^{-3}})$ increases forty-fold during the last 30 kyr, with negligible effects on its chemical composition. At moderate densities $(10^{10}<n_{\rm H}<10^{11}\ {\rm cm^{-3}})$ and cool temperatures $15<T<50$ K, increasing grain sizes delay molecular depletion. At low densities $(n_{\rm H}\sim10^{7}\ {\rm cm^{-3}})$, grains do not grow significantly. To assess the need to perform chemo-MHD calculations, we compared our results with a two-step model that reproduces well the abundances of C and O reservoirs, but not the N and S reservoirs.}
    {The chemical composition of the FHSC is heavily determined by that of the parent prestellar core. Chemo-MHD computations are needed for an accurate prediction of the abundances of the main N and S elemental reservoirs. The impact of grain growth in moderately dense areas delaying depletion permits the use of abundance ratios as grain growth proxies.}


    \titlerunning{Grain growth and its chemical impact in the First Hydrostatic Core phase}
    \maketitle

    \section{Introduction}
    
        The current picture of the star formation process identifies prestellar cores as the sites where stars are born. As revealed by the \emph{Herschel} space telescope \citep{Andre2010}, molecular clouds present filamentary structures with density enhancements that we call dense cores. These dense cores become prestellar cores as they undergo gravitational collapse when their density is high enough to become locally gravitationally unstable \citep{Jeans1902}. At this stage, the accreting gas radiates gravitational energy freely in an isothermal collapse. Once the gas becomes optically thick, the radiation is trapped, which heats the gas and forms the first Larson core \citep{Larson1969}. The first Larson core stage, also known as first hydrostatic core (FHSC) stage, is short-lived; it lasts up to $\sim 10^{4}$ yr, and is expected to end when the central region reaches temperatures of  $\sim 2000$ K and molecular hydrogen dissociates. This causes a second collapse, which brings the former FHSC to the typical size of a protostar and initiates the Class 0 phase. This is when the protostar is born.

        Given the short life and the low luminosity of FHSCs, these objects are difficult to detect. Their properties are instead inferred from numerical simulations of core collapse \citep[see, e.g.,][]{Masunaga1998, Masunaga2000, Price2009, Tomida2010, Commercon2012, Tomida2013, Bate2014, Hennebelle2016}. Apart from their short lives, of up to $\sim 10^{4}$ yr \citep{Boss1995}, the simulations describe FHSCs as faint sources \citep[$L_{\rm bol}<0.1 L_{\odot}$,][]{Bate2014} with temperatures of a few $10^{2}$ K, model-dependent sizes in the range of $5-20$ au, and hydrogen number densities higher than $10^{11}$ cm$^{-3}$ \citep{Hincelin2016}. Simulated FHSCs may also exhibit disks surrounding the object and low-velocity outflows \citep{Commercon2012, Hincelin2016}. Regarding the properties of the future star, the FHSC has recently been proposed to play a key role in setting the characteristic mass of stars, that is, the peak of the initial mass function \citep{Lee2018, Hennebelle2019}, which is expected to lie at about ten times the mass of the FHSC. The detection of molecular emission of refractory material and desorbed molecules in the outflow has helped identify and discard FHSC candidates \citep[see, e.g.,][]{Chen2010, Chen2012, Young2019, Maureira2020, Wakelam2022}. So far, only a few sources have been proposed as FHSCs attending to the properties found in numerical simulations. The embedded nature of FHSCs is apparent in the faint and compact infrared emission at 70 $\mu$m found in some candidates \citep[see, e.g.,][]{Evans2003, Enoch2010, Pineda2011}. The weak outflow activity predicted by the simulations is also present in some FHSC candidates \citep[see, e.g.,][]{Gerin2015}.

        Numerical simulations have therefore become the tool of choice to study, in a realistic manner, the dynamical evolution of this elusive stage of the star formation process. This dynamical evolution has been investigated following different approaches: analytical one-dimensional models \citep[see, e.g.,][]{Priestley2018, Bhandare2018}, two-dimensional semi-analytical models \citep{Visser2009, Drozdovskaya2014, Drozdovskaya2016}, and three-dimensional radiation-magnetohydrodynamic (R-MHD) simulations \citep[see, e.g.,][]{Commercon2012, Hincelin2016, Hennebelle2020}, both in its ideal \citep[][]{Teyssier2006} and non-ideal implementations \citep[see, e.g.,][]{Masson2012, Masson2016}. MHD simulations have become the state-of-the-art approach for the analysis of the gravitational collapse of prestellar cores into protostellar objects. The decisive role of the magnetic field in the process of collapse, fragmentation, and disk formation \citep{Allen2003, Hennebelle2008b, Commercon2010} make these tools essential to that end. Non-ideal MHD effects have a critical importance in cloud collapse, disk formation, and grain growth \citep{Li2014, Masson2016, Hennebelle2020, Gong2020, Gong2021, Lebreuilly2023}. In the formation of disks, the so-called magnetic braking catastrophe that leads to large, unrealistic disks \citep{Allen2003, Hennebelle2008a} is prevented by the introduction of diffusive processes such as the ambipolar diffusion, Ohmic dissipation, and the Hall effect \citep{Nakano2002}. Additionally, ambipolar diffusion has been shown to effectively remove small grains, with sizes of 0.1 $\mu$m and below, from the grain-size distribution, even in turbulent scenarios \citep[see, e.g.,][]{Lebreuilly2023}.

        Being the earliest stage in the formation of stars, the chemical characterization of FHSCs is of critical importance to understanding the chemical composition of the future protoplanetary disk and planetary system. One way to investigate the chemical composition of collapsing cores may involve the use of chemical models on the evolution of physical properties provided by the numerical simulation of collapse \citep{Hincelin2016}. On the one hand, according to the simulations, FHSC chemistry resembles a dark cloud chemistry in its envelope, with cold gas $T_{\rm kin}<20$ K and densities varying between $5\times 10^{4}$ cm$^{-3}$ and $10^{7}$ cm$^{-3}$ \citep{Hincelin2016}. In this scenario, the depletion of molecules (e.g., CO) and chemical reactions over dust grain surfaces are expected to occur, leading to a rich chemistry of interstellar complex organic molecules (iCOMs). The depletion of molecules on grain surfaces and the subsequent ice growth also change the grain size distribution, and have an important impact on the chemical processes dependent on the adsorption surface, the thermodynamics of dust, and processes such as the cosmic-ray induced desorption \citep{Iqbal2018}. These aspects, in the end, greatly affect the chemistry of these objects \citep[see, e.g.,][]{Sipila2020}. On the other hand, the desorption of molecules and gas-phase chemistry dominates the collapsing core chemistry once the temperature at the core reaches a few $\sim 100$ K, releasing the processed ice content locked on grain surfaces into the gas phase \citep{Hincelin2016}.

        Dust grains and dust grain sizes are therefore key actors in the star formation process and its chemistry. In the thermodynamics of molecular clouds, dust grains are involved in the synthesis of molecules that radiatively cool the gas \citep{Draine2011} and, at high densities $n>10^{5}$ cm$^{-3}$, dust dominates gas cooling \citep{Goldsmith2001}. Dust is also responsible for the far-ultraviolet (FUV) interstellar radiation extinction that allows the survival of molecules and, as noted above, dust is also involved in the synthesis of complex organic molecules on their surfaces. Over recent years it has become increasingly evident that dust size evolves during star formation. There is evidence of grain growth in dense molecular clouds, revealed by the study of extended dust emission known as coreshine. Dust continuum observations at 3.6 $\mu$m with the Spitzer Infrared Array Camera (IRAC) instrument revealed emission patterns that are only possible by large, micron-sized grain scattering of radiation \citep{Pagani2010, Steinacker2010}. Further observations at 3.6 and 4.5 $\mu$m also helped constrain grain sizes \citep{Steinacker2015} and, at 8 $\mu$m, \citet{Lefevre2016} showed that uncoagulated grains are unable to reproduce the observations. The processes of dust coagulation and fragmentation, the consequent changes in the grain size distribution, and their impact on the star formation process have been investigated with theoretical models of increasing complexity \citep[see, e.g.,][]{Ossenkopf1993, Ormel2009, Hirashita2009, Guillet2020, Silsbee2020, Marchand2021, Tsukamoto2021, Marchand2022, Tu2022, Marchand2023, Kawasaki2023, Lebreuilly2023}, taking into account the thermal motion of grains, turbulence, Brownian motion, and diffusive effects, namely ambipolar and Ohmic diffusion, and the Hall effect. These studies have shown a strong influence of grain growth in dust opacities and magnetic resistivities, which in turn affect the dynamics of collapsing cores and the formation of disks \citep[see, e.g.,][]{Marchand2023}.

        In this paper we present a study on the impact of dynamics and grain growth on the chemical composition of a FHSC. This study was carried out by postprocessing tracer particles from a MHD simulation \citep{Hennebelle2016} performed with the code \texttt{RAMSES} \citep{Teyssier2002}. The chemical postprocessing was done with the gas-grain chemical model \texttt{NAUTILUS} \citep{Ruaud2016}. We also investigated and discuss the effects of grain growth on the chemical composition of the FHSC with the code \texttt{SHARK}\footnote{\href{https://github.com/ulebreui/shark}{https://github.com/ulebreui/shark}} \citep{Lebreuilly2023}. In Sects. \ref{sec:physicalSimulations} and \ref{sec:chemModel} we describe the codes employed to carry out the simulation, grain growth, and the chemical modeling. In Sect. \ref{sec:results} we present the results of the chemical modeling, and we discuss the relevant chemical processes taking place at the different scales that this young stellar object (YSO) covers, the spatial distribution of chemical abundances, and their evolution through time. In Sect. \ref{sec:discussion} we discuss several aspects of the results of the chemical modeling: the chemical production and inheritance from the dense core to the hot core phase, the suitability of using simple two-step models to chemically describe a FHSC and, finally, we discuss the role of grain size in the chemical makeup of the FHSC.   
        
    \section{Physical simulations}\label{sec:physicalSimulations}
    
        The time evolution of physical properties of the collapsing core was computed by the adaptive mesh refinement (AMR) code \texttt{RAMSES} \citep{Teyssier2002}. Additionally, tracer particles introduced in this simulation followed the time evolution of physical quantities (density, temperature, and magnetic field) over time. They were then postprocessed with the code \texttt{SHARK} \citep{Lebreuilly2023} to study grain growth during core collapse.
        
        \subsection{The adaptive mesh refinement code \texttt{RAMSES} and non-ideal magnetohydrodynamics}
        
\begin{figure*}
                \centering
                \includegraphics[width = \textwidth]{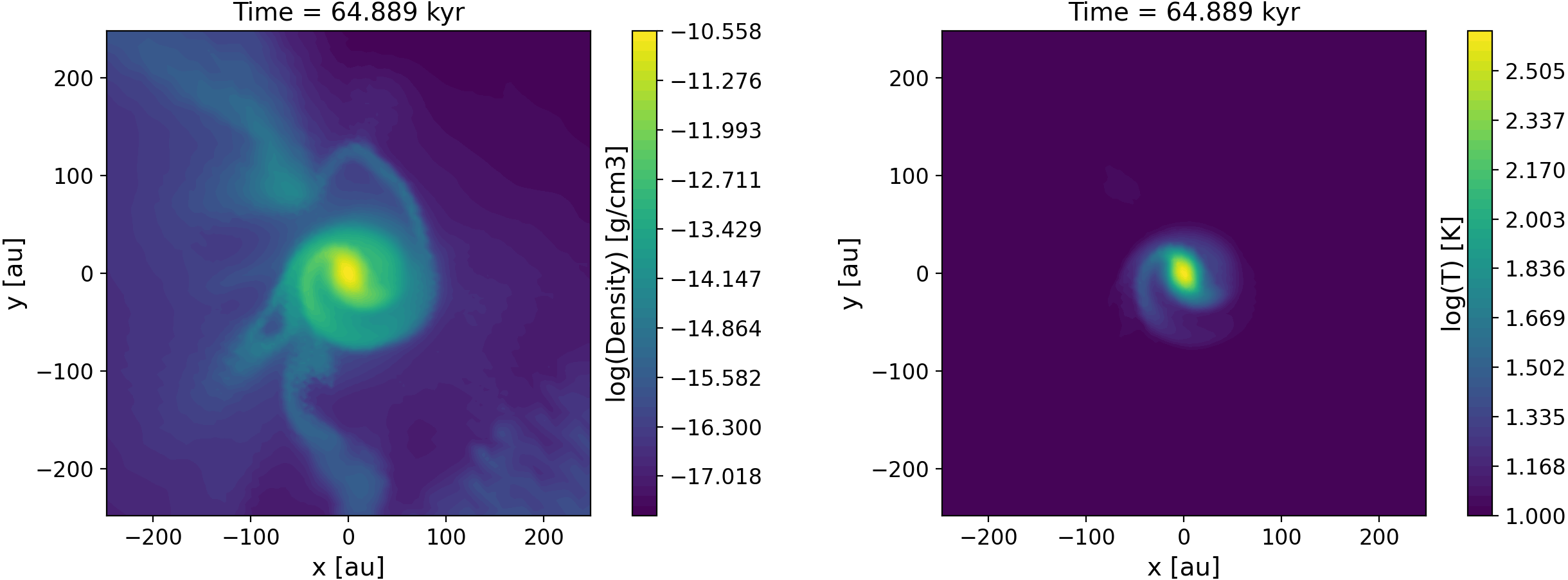}
                \caption{Density (left) and temperature (right) slices of the mesh through its origin in a face-on orientation at the end of the simulation ($6.5\times 10^{4}$ years).}
                \label{fig:lastSnapshot}
        \end{figure*}

            For the computation of the dynamical evolution of the collapsing core we used the MHD solver of the AMR code \texttt{RAMSES} in its non-ideal MHD implementation \citep{Masson2012}. We analyzed the simulation from \citet{Hennebelle2016} and \citet{Gerin2017}, in which only ambipolar diffusion, following the prescription of \citet{Marchand2016}, is considered as non-ideal MHD process. The initial setting of our simulation is a 1 $M_{\odot}$ strongly magnetized (initial mass-to-flux ratio of $\mu=2$) and turbulent (velocity fluctuations match an initial Mach number $\mathcal{M}=1.2$) sphere of radius $r\sim 2500$ au, uniform density of $\rho= 10^{-17}$ g cm$^{-3}$, and uniform temperature of 10 K. The barotropic equation of state governs the interplay between temperature and density. This equation of state works as an efficient approximation to a more accurate radiative transfer of the FHSC provided by, for example, the hybrid radiation scheme presented in \citet{Mignon2020}, as long as no protostar is formed, which is our case. The mesh is refined to always describe the Jeans length with at least 12 points, down to a resolution of $\sim 0.15$ au. The physical evolution of the collapsing core is computed until the FHSC phase is reached in $6.5\times 10^{4}$ years. In Fig. \ref{fig:lastSnapshot} we show the density and temperature maps of the collapsing core at this stage.

        \subsection{Tracer particles}

            To probe the physical properties of the collapsing fluid and follow their evolution we introduced tracer particles in the \texttt{RAMSES} code. Each tracer particle provides, among other physical quantities, the density and temperature evolution with time. In this simulation, $2\times 10^{5}$ tracer particles were set to provide this information. The time evolution of density and temperature probed by the tracer particles allowed us to compute the chemical evolution of the collapsing core using the chemical code \texttt{NAUTILUS}.

        \subsection{Grain growth with \texttt{SHARK}}

            Aiming to study the impact of the dust grain size on the chemistry of the FHSC, we used the \texttt{SHARK} code \citep{Lebreuilly2023}, a one-dimensional, spherically symmetric, and finite-volume code that solves the hydrodynamics of a mixture of gas and dust. We included dust growth as in \citet{Lebreuilly2023} in the low fragmentation threshold case. Dust is allowed to grow and fragment in grain-grain collisions induced by dust differential velocities caused by turbulence \citep{Ormel2007}, brownian motion, and ambipolar diffusion. The required energy for the fragmentation of silicates is taken from the theoretical considerations of \citet{Dominik1997} and the experimental results of \citet{Blum2000}. This code also calculates the charge of dust grains, the density of ions and electrons, and the Ohmic, ambipolar, and Hall resistivities. \texttt{SHARK} is a polyvalent code that allows the creation of different setups to meet the modeling features of the astrophysical objects under consideration, such as molecular clouds, prestellar/protostellar cores, photo-dissociative regions (PDRs), and protoplanetary disks.

    \section{Chemical modeling}\label{sec:chemModel}
        
        The chemical modeling of the simulation was performed as a postprocessing of the physical evolution provided by the $2\times 10^{5}$ tracer particles with the chemical code \texttt{NAUTILUS}. \texttt{NAUTILUS} \citep{Ruaud2016} is a three-phase chemical code in which gas, grain surface and mantle phases, and their interactions, are considered. Given an initial set of physical and chemical conditions, \texttt{NAUTILUS} computes the evolution of chemical abundances of 1126 chemical species through time. 

        \subsection{Initial conditions and chemical postprocessing}

            In this paper, the time evolution of temperature-density pairs provided by the tracer particles was given as physical input to compute chemical abundances. To take into account the deeply embedded nature of collapsing cores and the role of the parent cloud in photochemistry, we considered the physical evolution of the collapsing core taking place in an ambient cloud of 30 mag of visual extinction. Our initial setup consists of an evolved dense core whose chemical makeup is already very different from that of a diffuse cloud. More precisely, we included the chemical abundances after $10^{6}$ yr at the extinction peak of Barnard 1b obtained in \citet{NavarroAlmaida2020} as initial abundances for the simulation. These initial abundances were determined with a one-dimensional fitting to the observational data of the GEMS IRAM 30m Large Program (PI: Asunci\'on Fuente) toward Barnard 1b. We chose this source as it hosts extremely YSOs that are well suited for our study. Moreover, the simulation we postprocessed here was previously used in \citet{Gerin2017} to fit the spectral energy distribution and continuum ALMA data of the same source. After setting the initial abundances equally for all tracer particles present in the simulation, we took the evolution of density and temperature obtained with the tracer particles to compute the evolution of chemical abundances beyond the initial abundances set before. In this process, dust temperatures were considered to be equal to gas temperatures. The cosmic-ray ionization rate was set to $\zeta_{\rm H_{2}} = 6.5\times 10^{-17}$ s$^{-1}$ and the interstellar FUV field strength, in Draine field units, was set equal to $\chi_{\rm UV}= 24$, the values used in \citet{NavarroAlmaida2020}. While it is known that the cosmic-ray ionization rate decreases with the column density \citep[see, e.g.,][]{Caselli1998, Padovani2009}, in this version of \texttt{NAUTILUS}, these quantities cannot evolve with time or position, and therefore were kept constant. This may lead to an overestimation of gas-phase abundances in regions where cosmic-ray desorption or cosmic ray-induced UV photodesorption are relevant. Finally, for the grain surface chemistry, we adopted a chemical desorption scheme with reduced efficiency to account for the presence of icy grain surfaces \citep{Minissale2016}.
        
        \subsection{Chemical network}

            The chemical network we used in this analysis was the one presented in \citet{NavarroAlmaida2020}. This chemical network uses the data available in the KIDA database. It is composed of 1126 species (588 in the gas phase and 538 in solid phase) linked together via 13155 reactions. In addition, it includes an up-to-date sulfur chemical network including recent updates \citep{Fuente2017, LeGal2017, Vidal2017, LeGal2019}. Thermal and nonthermal desorption into the gas phase is only allowed for species on grain surfaces. The nonthermal desorption mechanisms present in the network include desorption induced by cosmic-rays \citep{Hasegawa1993}, photodesorption, and chemical desorption \citep{Garrod2007, Minissale2016}.
    
    \section{Results}\label{sec:results}

        The evolution of the physical structure provided by \texttt{RAMSES} was computed for $6.5\times 10^{4}$ years. We first describe the physical properties of the collapsing core at three different times corresponding to different stages: initial setting, dense and cold prestellar phase, and FHSC phase.

        \subsection{Physical properties of the collapsing core}
        
\begin{figure*}
                \centering
                \includegraphics[width = \textwidth]{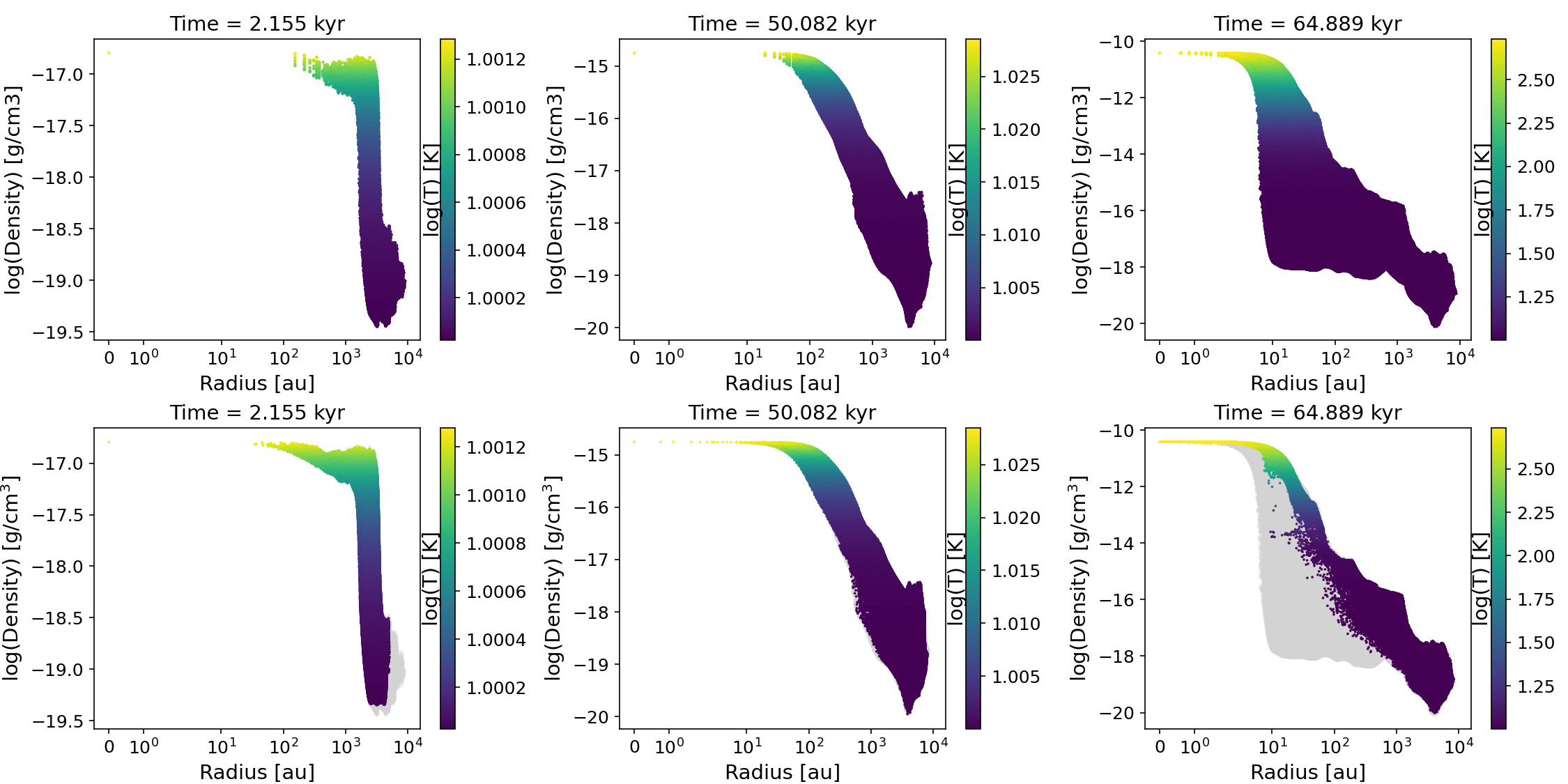}
                \caption{Comparison between the physical properties (density in the y-axis and temperature as a color gradient) and positions of the cells present in the \texttt{RAMSES} mesh (colored dots in the top row and gray areas in the bottom row) and those of the tracer particles (colored dots in the bottom row) at three different times, one for each column.}
                \label{fig:sampling}
            \end{figure*}

            In Fig. \ref{fig:sampling} we show the physical properties of the cells in the mesh of our simulation at three different times: $2-50-65\times 10^{3}$ years. As a consequence of the barotropic equation of state used to model the thermodynamics of the gas, density and temperature have a one-to-one relationship, although with different dynamic ranges. At $t=2\times 10^{3}$ yr, the beginning of our simulation, we are in a cold, moderately dense prestellar phase. Cell temperatures are mostly uniform at $\sim 10$ K, while density ranges from $10^{-20}$ g cm$^{-3}$ to $10^{-17}$ g cm$^{-3}$, that is, in the range $\sim 10^{4}-10^{7}$ cm$^{-3}$ in atomic hydrogen number density. As the collapse does not progress significantly at this stage, the mesh toward the center is not refined and therefore areas at a distance of up to $\sim 100$ au from the center are covered by a single cell centered at the origin (see left column of Fig. \ref{fig:sampling}). The collapse proceeds isothermally at 10 K until $\sim 5\times 10^{4}$ years. At this age, the maximum density toward the center increases two orders of magnitude while the temperature remains at $\sim 10$ K (middle column of Fig. \ref{fig:sampling}). Finally, entering the FHSC phase, the collapse becomes adiabatic and the released gravitational potential energy heats the protostar. At the end of our simulation, the density at the center increases up to $\sim 10^{-10}$ g cm$^{-3}$ while the temperature reaches a few $\sim 100$ K.

\begin{figure*}
                \centering
                \includegraphics[width = \textwidth]{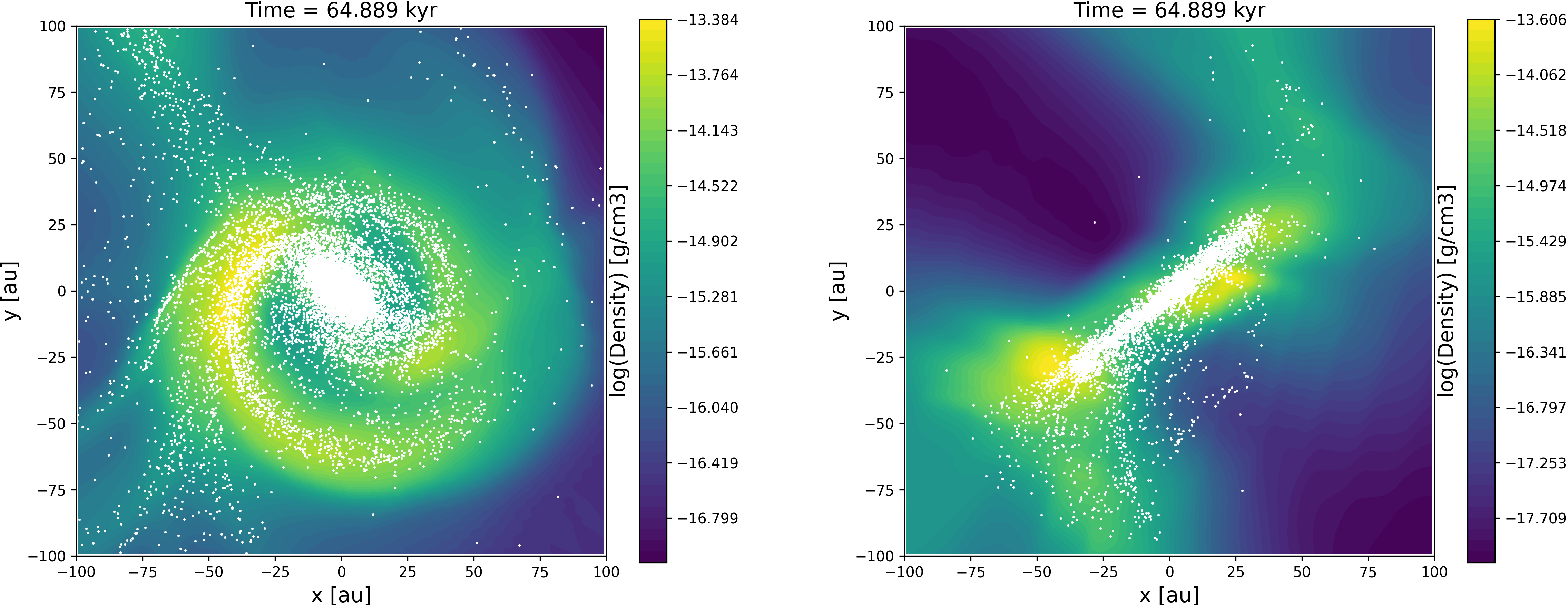}
                \caption{Spatial distribution of the tracer particles within a radius of 100 au from the center (white dots) overlapping column density maps at different viewing angles: face-on (left) and edge-on (right).}
                \label{fig:outflow}
            \end{figure*}

            In Fig. \ref{fig:sampling} we also show the density and temperature of the tracer particles at the snapshots chosen previously. The comparison between the density and temperature of the cells and those of the tracer particles offers information of the areas that are best sampled. While in the initial and intermediate stages the tracer particles provide a good sampling of the physical conditions present in the mesh (Fig. \ref{fig:sampling}), at the end of the simulation there is a significant lack of tracer particles toward the low and moderate density cells at a distance of $10-100$ au from the center. This region would correspond to the outflow cavity of the FHSC. In the simulations performed by \citet{Hincelin2016}, the outflow is the less populated component of the FHSC by tracer particles. This is apparent in Fig. \ref{fig:outflow}, where only few tracer particles are present in the outflow cavity of the FHSC. The plane of rotation in the inner 20 au is however well sampled (see Fig. \ref{fig:outflow}).

        \subsection{Distribution of chemical abundances: Temperature components}\label{sec:tempComponents}

            The tracer particles present in this simulation allowed us to follow the chemical evolution throughout the collapse of the prestellar core until the FHSC phase. Each tracer particle provided an evolution of temperature and density with time that is then processed with \texttt{NAUTILUS}, obtaining the spatial distribution and time evolution of chemical abundances across the protostar. For the chemical characterization of the FHSC, we postprocessed the $2\times 10^{5}$ tracer particle histories with \texttt{NAUTILUS} using a single grain size of $\bar{r} = 10^{-5}$ cm. The discussion of the impact of this parameter in the chemistry of the FHSC is carried out in Sect. \ref{sec:grainGrowth}.

            \begin{figure*}
                \centering
                \includegraphics[width = \textwidth]{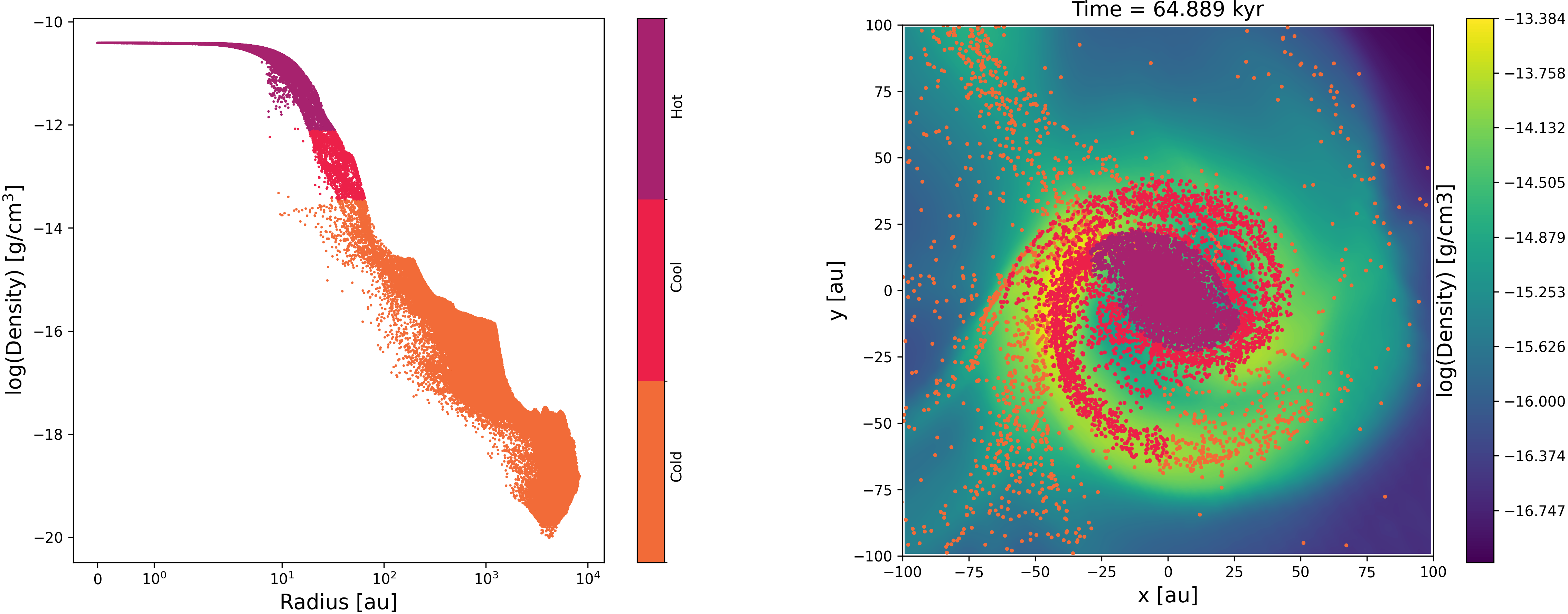}
                \caption{Definition of the temperature components described in Sect. 4.2. \emph{Left:} Values of distance and density of the temperature components. \emph{Right:} Spatial distribution of the temperature components with the same color-coding as in the left panel, overlapped to the face-on density map of the FHSC.}
                \label{fig:tempComponents}
            \end{figure*}

            To analyze the spatial distribution of chemical abundances, we defined sets of tracer particles according to their temperature at the end of the simulation (see Fig. \ref{fig:tempComponents}). As shown there, these temperature components trace different spatial scales, in which various chemical processes occur. In cold and dense gas, adsorption of molecules and ice growth are expected to be relevant chemical processes. Since CO is one of the most abundant molecules in interstellar ices, the cold component of the gas was defined as the set of tracer particles such that their temperature at the end of the simulation is up to 15 K (50\% of the total of tracer particles), when CO thermal desorption is triggered at moderate densities $\sim 10^{5}$ cm$^{-3}$ typical of molecular clouds \citep{MunozCaro2010}. As it can be seen in Figs. \ref{fig:sampling} and \ref{fig:tempComponents}, the dynamic range of the temperature in this component is very low compared to that of the density as a result of the barotropic equation of state. Consequently, the chemical differences that may appear in this component are expected to be primarily influenced by density. Another temperature component, the cool component, was defined as the set of tracer particles whose temperature lies in the range $15-50$ K (2\% of the total). Finally, the hot component of the gas collects the remaining tracer particles (48\% of the total) with temperatures up to a few $100$ K at the end of the simulation. Given the hot and compact nature of this component at the end of the simulation (see Fig. \ref{fig:tempComponents}), it can be considered as an incipient hot corino. The molecular abundances obtained in the following sections, rich in iCOMs and products of ice mantle sublimation, support this denomination. Unlike the cold component, in this hot component the dynamic range of the temperature is larger, reaching from $50$ K to several $100$ K when density increases a factor of $\sim 100$. The role of the temperature in this component is therefore expected to be more important.        

\begin{figure*}
                \centering
                \includegraphics[width = \textwidth]{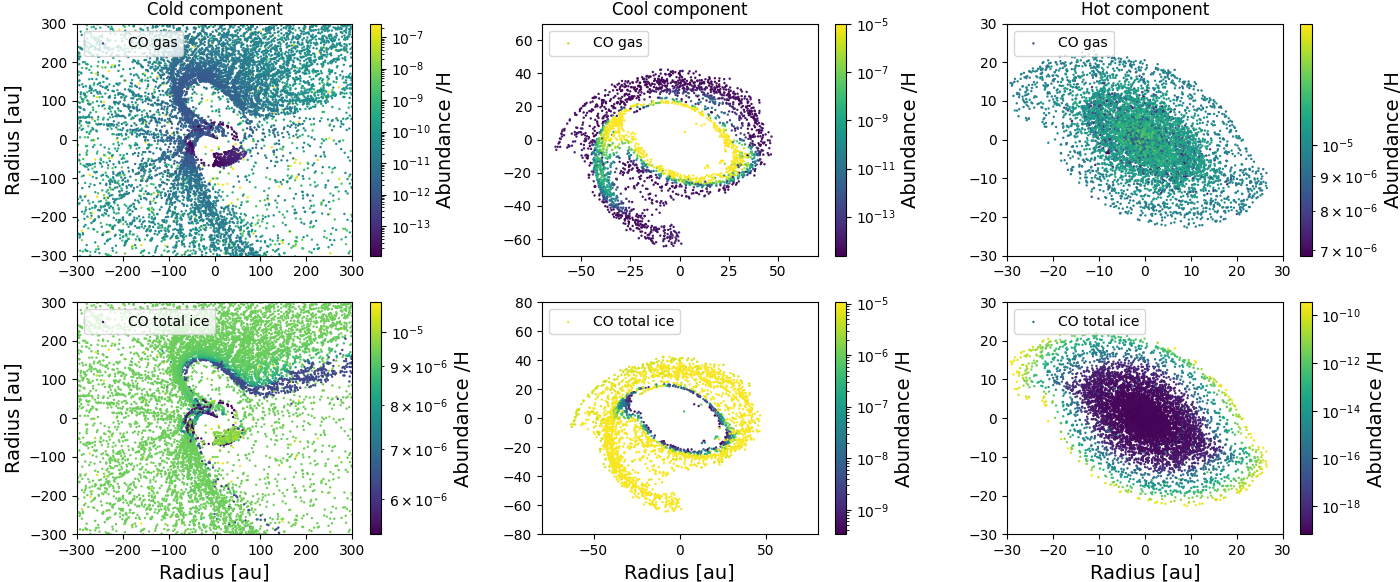}
                \caption{Spatial distribution of the tracer particles at the end of the simulation and their CO gas-phase (top row) and total ice-phase (bottom row) abundances  for the three temperature components: cold (left), cool (middle), and hot (right).}
                \label{fig:componentsCO}
            \end{figure*}

            In Fig. \ref{fig:componentsCO} we show the CO gas and ice-phase abundances at the end of the simulation for the three temperature components. The details of the distribution of gas-phase CO abundance are summarized in Table \ref{tab:distributionCOGas}. In Appendix \ref{sec:appendixHistograms}, the statistical distribution of gas-phase abundances for the rest of molecules is provided. While the temperature of the cold component is mostly uniform, its density increases up to six orders of magnitude. In such environment, the adsorption of molecules onto grain surfaces is expected to vary greatly. This is precisely the behavior displayed in Fig. \ref{fig:componentsCO}, where the adsorption of gas-phase CO molecules onto the grains increases toward the center, enhancing the total CO ice abundance. The gas-phase abundance of CO in this component spans seven orders of magnitude, in the range $10^{-14}-10^{-7}$. The cool component shows the transition from a scenario of heavy molecular depletion to another scenario in which molecules trapped in the ice matrix start to thermally desorb. In this component we observe a great dispersion of the CO gas-phase abundance, spanning ten orders of magnitude $10^{-15}-10^{-5}$. This results in thermal desorption of icy CO molecules toward the center. Finally, in the hot component we expect high gas-phase abundances of most of the molecules due to thermal desorption, with little dispersion. The gas-phase abundance of CO is in the range $(0.7-1.5)\times 10^{-5}$. The icy CO abundances are, consequently, highly diminished in this component.
        
            \begin{table}
                \centering
                \caption{Distribution (maximum, minimum, first and third quartiles, and median) of CO gas abundance in the temperature components.}
                \resizebox{0.495\textwidth}{!}{
                \begin{tabular}{ccccc}
                    \toprule
                    & & Cold component & Cool component & Hot component \\
				    \midrule\midrule
                    & Minimum   & $1.15\times10^{-14}$  & $1.06\times10^{-15}$  & $6.86\times10^{-6}$ \\
                    & Maximum   & $2.70\times10^{-7}$   & $2.39\times10^{-5}$   & $1.51\times10^{-5}$ \\
				CO  & Q$_{1}$   & $2.74\times10^{-10}$  & $5.77\times10^{-15}$  & $1.03\times10^{-5}$ \\ 
                    & Q$_{3}$   & $1.34\times10^{-7}$   & $4.34\times10^{-7}$   & $1.24\times10^{-5}$ \\
                    & Median    & $7.81\times10^{-9}$   & $8.44\times10^{-13}$  & $1.14\times10^{-5}$ \\
                    \midrule
                    \bottomrule
                \end{tabular}
                }
                \label{tab:distributionCOGas}
            \end{table}

            We conclude that gas density is a critical property for the chemistry of the cold component, dominated by the adsorption of molecules due to its cold temperature $10-15$ K and moderate density. In the cool component, the thermal desorption temperature of the molecular species determines the chemical abundances. A large portion of ice is lost here due to the thermal desorption of CO. Molecules such as H$_{2}$O or iCOMs, with desorption temperatures of $\sim 100$ K, are not desorbed effectively in this component. Finally, the hot component is dominated by the thermal desorption of ices and the enhancement of gas-phase abundances, with the exception of the ions included in our analysis, that is, N$_{2}$H$^{+}$ and HCO$^{+}$. We performed a similar analysis for the remaining molecules considered in this paper. The spatial distribution of their abundances and the details of such distributions is shown in Appendix \ref{sec:spatialDist}. So far we have examined the chemical composition at different spatial scales of the final snapshot of the MHD simulation. In the following section we analyze the time dependence of chemical abundances in the different temperature components previously defined.

        \subsection{Evolution of chemical abundances}

\begin{figure*}
                \centering
                \includegraphics[width = \textwidth]{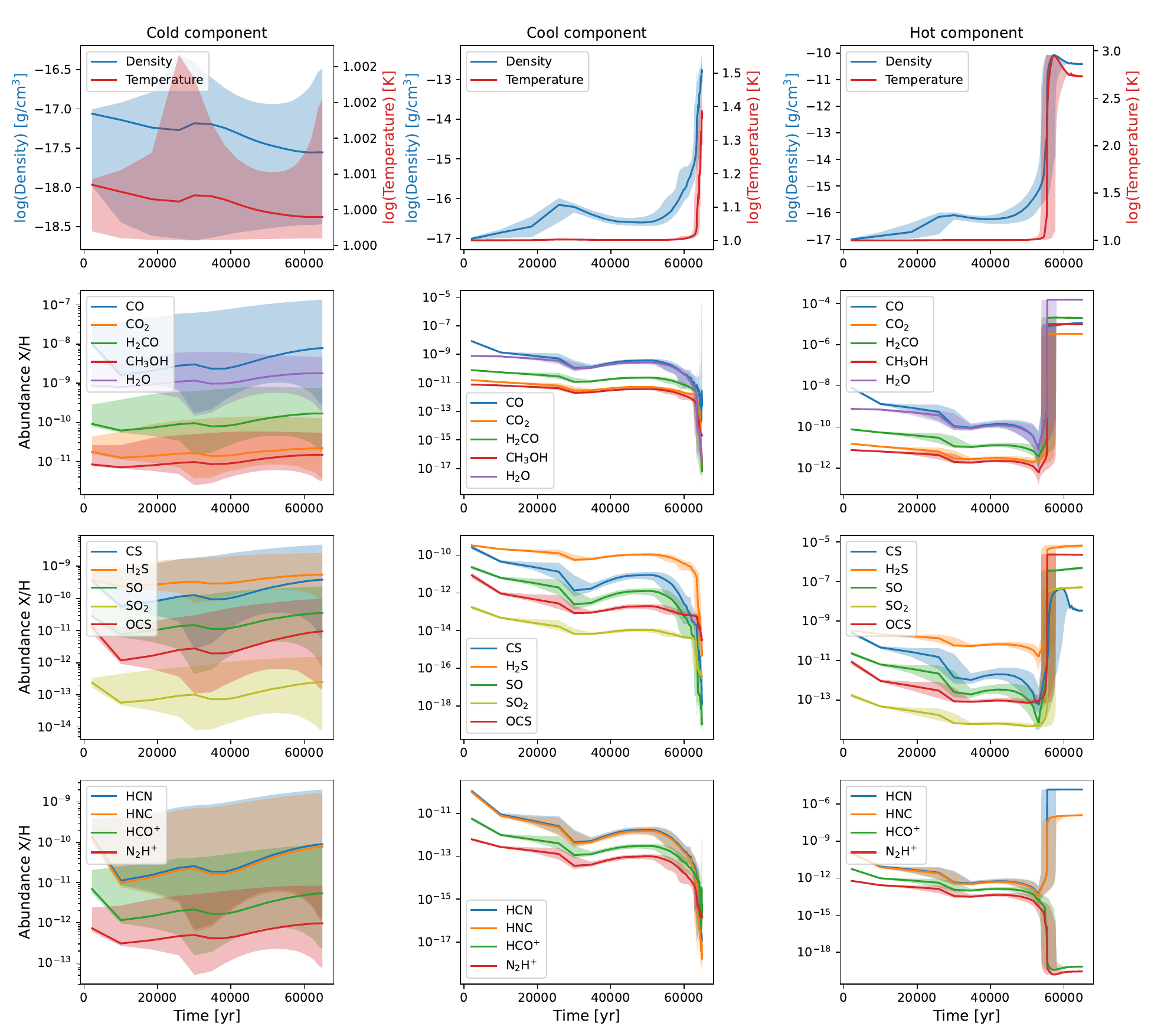}
                \caption{Time evolution of the density and temperature (top row) and gas-phase abundances (remaining rows) during the core collapse for the three temperature components: cold (left), cool (middle), and hot (right).}
                \label{fig:timeEvolutionAbs}
            \end{figure*}

            At any given snapshot, the set of $2\times 10^{5}$ tracer particles probe a wide variety of physical conditions depending on the spatial scales they cover. Similarly, the same variety of physical properties can be explored by the time evolution of tracer particles throughout the simulation. Since molecular abundances are mainly determined by evolving physical properties, we expect to find similarities between the radial dependence of chemical features and its time dependence. In this section we perform a statistical analysis of the time evolution of chemical abundances in each temperature component. The cold component is characterized by a mostly uniform temperature $\sim 10$ K and an ample range of densities. The time evolution of density and temperature is shown in Fig. \ref{fig:timeEvolutionAbs}. The median values of density and temperature decrease over time, with an increasing dispersion. This behavior allows gas-phase abundances to increase their median values as they suffer from decreasing depletion rates. Gas-phase abundance distributions also become wider as a consequence of the increasing dispersion of both density and temperature in this component. This behavior is present in the gas-phase abundances of the carbon, oxygen, sulfur, and nitrogen carriers present in Fig. \ref{fig:timeEvolutionAbs}. Finally, a constant abundance ratio HCN/HNC of 1 is measured at all times, which reflects the cold nature of this component \citep{Hacar2020, Navarro2023}.
            
            The cool component is denser than the cold component, with a narrow and increasing density range as seen in Fig. \ref{fig:timeEvolutionAbs}. Its temperature ranges from 15 to $\sim 50$ K at the end of the simulation. Unless the species is thermally desorbed, higher densities lead to a higher depletion of gas-phase abundances, seen in Fig. \ref{fig:timeEvolutionAbs}. Depletion rates however greatly vary among species. We defined a depletion rate as the inverse of the time elapsed between the local maximum in gas-phase abundances at $\sim 5\times 10^{4}$ yr and the time instant in which the abundance drops to a $10\%$ of the value at the local maxima previously defined. We identified two families of molecules: fast depleting and slow depleting ones (see Table \ref{tab:depletionTimes}). Molecules such as CS, CO, H$_{2}$O, HCN, HNC, or SO show a faster depletion than other molecules such as SO$_{2}$, H$_{2}$S, H$_{2}$CO, OCS, or CH$_{3}$OH. Moreover, the slow depleting molecules in this sample often possess routes of formation involving reactions over grain surfaces and chemical desorption. This would make their gas-phase abundance more resilient to depletion than the rest of molecules. Toward the end of the simulation, temperatures from 15 K to 50 K are reached. Since the CO desorption temperature was included in this range, we detected a widening of the CO gas-phase abundance shown in Fig. \ref{fig:timeEvolutionAbs} toward the end of the simulation.

            \begin{table}
                \centering
                \caption{$X_{\rm 50kyr}$, the gas-phase abundances at $t\sim 50$ kyr; $t_{10\%}$, the time needed to reduce them by one order of magnitude; and its inverse, the depletion rate.}
                \resizebox{0.495\textwidth}{!}{
                \begin{tabular}{lccc}
                    \toprule
                    Molecule & Abundance $X_{\rm 50kyr}$ & Depletion time $t_{10\%}$ (yr) & Depletion rate $t_{10\%}^{-1}$ (yr$^{-1}$)\\
				    \midrule\midrule
                    CO & $3.72\times 10^{-10}$ & $1.05\times 10^{4}$ & $9.52\times 10^{-5}$\\
                    CO$_{2}$ & $5.15\times 10^{-12}$ & $1.35\times 10^{4}$ & $7.41\times 10^{-5}$\\
                    H$_{2}$O & $2.69\times 10^{-10}$ & $1.07\times 10^{4}$ & $9.35\times 10^{-5}$\\
                    H$_{2}$CO & $2.34\times 10^{-11}$ & $1.31\times 10^{4}$ & $7.63\times 10^{-5}$\\
                    CH$_{3}$OH & $3.69\times 10^{-12}$ & $1.32\times 10^{4}$ & $7.58\times 10^{-5}$\\
                    CS & $9.09\times 10^{-12}$ & $8.45\times 10^{3}$ & $1.18\times 10^{-4}$\\
                    H$_{2}$S & $1.11\times 10^{-10}$ & $1.29\times 10^{4}$ & $7.75\times 10^{-5}$\\
                    SO & $1.27\times 10^{-12}$ & $8.84\times 10^{3}$ & $1.13\times 10^{-4}$\\
                    SO$_{2}$ & $1.08\times 10^{-14}$ & $1.33\times 10^{4}$ & $7.52\times 10^{-5}$\\
                    OCS & $1.99\times 10^{-13}$ & $1.37\times 10^{4}$ & $7.30\times 10^{-5}$\\
                    HCN & $1.72\times 10^{-12}$ & $1.00\times 10^{4}$ & $1.00\times 10^{-4}$\\
                    HNC & $1.51\times 10^{-12}$ & $9.87\times 10^{3}$ & $1.01\times 10^{-4}$\\
                    HCO$^{+}$ & $3.02\times 10^{-13}$ & $1.14\times 10^{4}$ & $8.77\times 10^{-5}$\\
                    N$_{2}$H$^{+}$ & $9.93\times 10^{-14}$ & $1.11\times 10^{4}$ & $9.01\times 10^{-5}$\\
                     \midrule
                    \bottomrule
                \end{tabular}
                }
                \label{tab:depletionTimes}
            \end{table}
            
            Finally, before the ignition of the hot core, the high density and relatively cold temperatures at the beginning of the simulation lead to depletion of gas-phase abundances in a behavior very similar to that found in the cool component (see Fig. \ref{fig:timeEvolutionAbs}). Once the temperature increases up to several $\sim 100$ K, the ice content is sublimated, enhancing gas-phase abundances. However, molecular ions such as HCO$^{+}$ and N$_{2}$H$^{+}$ are destroyed by their reaction with gaseous water \citep{Bergin1998} and CO \citep{Busquet2011}, respectively. The final gas-phase abundances of the hot component are the product of the thermal desorption of ice mantles, coming either from the depleted species during the simulation or the ice mantles inherited from the dense core phase calculated by \citet{NavarroAlmaida2020}, and additional gas and grain-phase chemical processes taking place during the heat up phase of the FHSC. The comparison between the initial abundances and the chemical abundances on the hot component is discussed in the next section in the context of chemical inheritance from prestellar cores to more evolved YSOs. The lower gas-phase CO abundance at the end of the simulation, compared to the canonical dense cloud abundance of $\sim 10^{-4}$ \citep{Lacy1994}, indicates a chemical processing of CO during the prestellar and protostellar collapse phases. CO is processed into organic molecules such as H$_{2}$CO and CH$_{3}$OH, whose abundance is similar or even higher than that of CO. For instance, the H$_{2}$CO/CO ratio before the ignition of the hot core is $\sim 0.1$, while it becomes $\sim 2$ when temperatures of a few $\sim 100$ K are reached, signaling to a high abundance of H$_{2}$CO ice. The gas-phase abundance of CH$_{3}$OH follows a similar behavior, with a ratio of CH$_{3}$OH/CO $\sim 0.01$ before the ignition of the protostar that then increases up to $\sim 1$ when the ice content is sublimated. Again, this indicates a high abundance of solid methanol locked in ices as a result of CO processing during both the prestellar and protostellar collapse phases. The hot nature of this component is consistent with the high HCN/HNC ratio, of $\sim 100$, at the end of the simulation. It is worth noting the low dispersion of the gas-phase abundance distributions at the end of the simulation. This means that, from the chemical point of view, a single tracer particle can be selected as representative of the hot component.
    
    \section{Chemistry on the Solar System scale}\label{sec:discussion}

        As defined in Sect. \ref{sec:tempComponents}, the hot component is comprised by the tracer particles whose temperature at the end of the simulation is higher than 50 K. The distribution of tracer particles in this set covers a region of radius $\sim 30$ au from the center (see Fig. \ref{fig:componentsCO}). This region is of particular interest as it is the most efficient planet-forming region in the future protoplanetary disk. The chemical analysis of this component is therefore of crucial importance to understanding the chemistry of more evolved YSOs.
    
        \subsection{Chemical reservoirs}\label{sec:reservoirs}

            \begin{figure*}
                \centering
                \includegraphics[width = \textwidth]{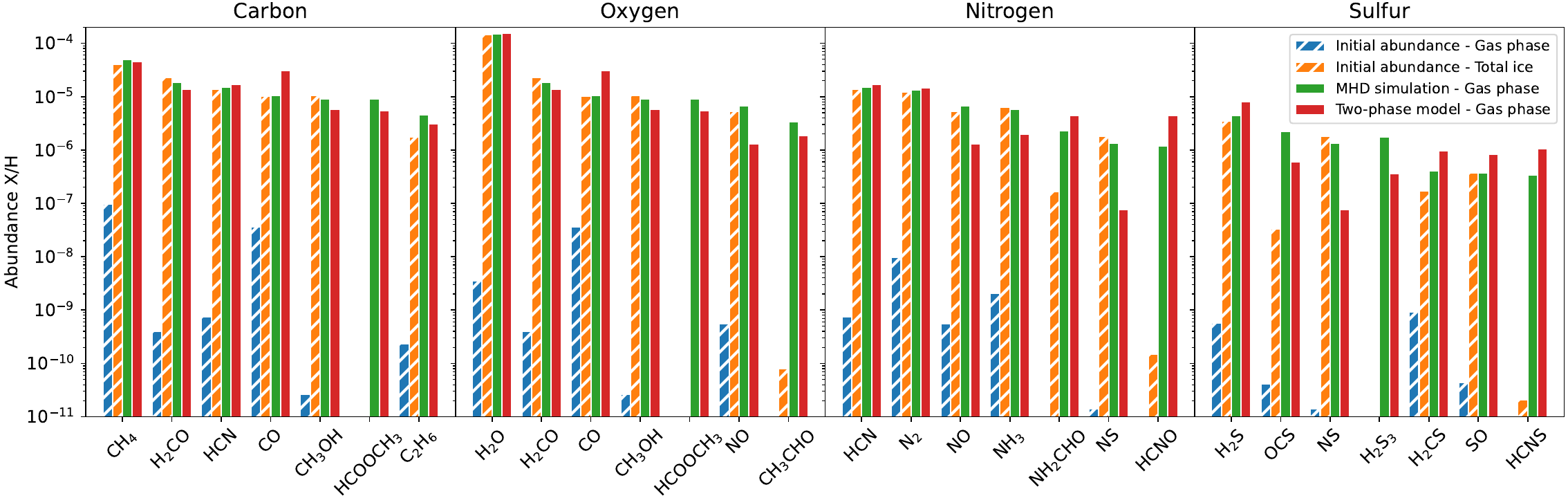}
                \caption{Comparison between the chemical abundances of the main elemental reservoirs in the hot component in four different settings: the initial gas-phase and ice-phase abundances (hatched blue and orange bars, respectively; see Sect. \ref{sec:chemModel}), the final gas-phase abundances predicted by the MHD simulation (green bars), and the final gas-phase abundance of the two-step model described in Sect. \ref{sec:twoPhaseModel} (red bars).}
                \label{fig:twoPhaseModel}
            \end{figure*}
        
            As discussed in Sect. \ref{sec:results} and shown in Fig. \ref{fig:histograms}, the distribution of chemical abundances of the hot component at the end of the simulation is highly centered toward the median, showing very low dispersion. This behavior offered the possibility of using one tracer particle belonging to this component to perform a detailed chemical analysis of the complete set of chemical species present in the chemical network used with the code \texttt{NAUTILUS}. Our first analysis collects the main reservoirs of carbon, oxygen, nitrogen, and sulfur in the hot component at the end of the simulation (see green bars in Fig. \ref{fig:twoPhaseModel}).
            
        \subsubsection{Carbon reservoir}
            
            According to our model, the most abundant carbon-bearing molecules in the hot component of the FHSC are CH$_{4}$, H$_{2}$CO, HCN, CO, CH$_{3}$OH, HCOOCH$_{3}$, and C$_{2}$H$_{6}$ (Fig. \ref{fig:twoPhaseModel}). Methane (CH$_{4}$) as the main carbon reservoir was also predicted by \citet{Hincelin2016}. The hydrogenation of CO is responsible for the high relative gas-phase abundance of organics such as H$_{2}$CO, with H$_{2}$CO/CO $\sim 1.80$, and CH$_{3}$OH, with CH$_{3}$OH/CO $\sim 0.87$. The chemical features predicted here have been measured in subsequent stages of the star formation process. For instance, the molecular complexity and high gas-phase abundances of formaldehyde (H$_{2}$CO), methanol (CH$_{3}$OH), and methyl formate (HCOOCH$_{3}$) are detected in line surveys of hot cores and corinos \citep[see, e.g.,][]{Jorgensen2016, Pagani2017}, when processed ices are sublimated as the protostar reaches temperatures of a few $\sim 100$ K. As a consequence of this chemical processing into iCOMs, the gas-phase abundance of CO in our simulation is found to be $1.17\times 10^{-5}$, an order of magnitude below the canonical interstellar dense cloud value of $\sim 10^{-4}$ \citep{Lacy1994}. This lack of CO is routinely measured toward Class 0 protostars \citep[see, for instance,][]{Yidiz2010, Anderl2016} and protoplanetary disks \citep[see, e.g.,][]{Albi2010, Bruderer2012, Albi2018}. The chemical processing of CO into iCOMs is invoked in order to explain this absence \citep{Bergin2014, Yu2016, Bosman2018}.

        \subsubsection{Oxygen reservoir}
            
            Following the same trend found in the set of main carbon reservoirs, organic molecules are among the most abundant oxygen-bearing molecules. In particular, these molecules are H$_{2}$O, H$_{2}$CO, CO, CH$_{3}$OH, HCOOCH$_{3}$, NO, and CH$_{3}$CHO. Our estimate for the gas-phase abundance of H$_{2}$O in the hot component at the end of the simulation, of $1.52\times 10^{-4}$, is compatible with the water ice content measured toward YSOs using 3 $\mu$m absorption spectroscopy \citep[$\sim 10^{-4}$, see, e.g.,][]{Pontoppidan2004}. Carbon monoxide, complex organic molecules formed by hydrogenation of CO, and NO complete the set of the most abundant reservoirs of oxygen.
                
        \subsubsection{Nitrogen reservoir}
            
            The main nitrogen reservoirs are, according to our model, HCN, N$_{2}$, NO, NH$_{3}$, NH$_{2}$CHO, NS, and HCNO. The gas-phase abundance of HCN becomes the highest among the nitrogen-bearing molecules. Owing to the high temperatures of the hot component and the temperature-dependent routes of formation of HNC, its abundance is two orders of magnitude below that of HCN. N$_{2}$, previously thought as the main reservoir of nitrogen \citep[see, e.g.,][]{Womack1992}, has a gas-phase abundance slightly below that of HCN due to the modifications in reaction rates involving the formation of N$_{2}$ in dense clouds \citep{Daranlot2012} that reduce its gas-phase abundance. As predicted by \citet{Herbst1986}, nitric oxide (NO) is a highly abundant nitrogen-bearing molecule after N$_{2}$. Formamide (NH$_{2}$CHO) is a complex organic molecule that has been identified as precursor of a large variety of prebiotic molecules \citep{Sepulcre2019}. Amino acids and proteins are built upon the same peptide bonds this molecule possesses. Formamide has been detected in plenty of low and intermediate-mass protostellar cores hosting hot cores and corinos \citep[see, e.g.,][]{Kahane2013, LopezSepulcre2015, Marcelino2018, Bianchi2019} and several works have investigated its formation in protostellar environments via gas \citep{Barone2015} and grain-phase \citep{Garrod2008, Song2016, Rimola2018} reactions. HCNO, another reservoir of nitrogen, is indeed proposed as a precursor of formamide in interstellar ices, where it becomes formamide after hydrogenation. Nitrogen sulfide, NS, is an abundant nitrogen and sulfur-bearing radical that has been used in studies of sulfur depletion. More precisely, the NS/N$_{2}$H$^{+}$ abundance ratio has been proposed as a quantity that provides a direct constrain on the abundance of gas-phase atomic sulfur \citep{HilyBlant2022}.
                
        \subsubsection{Sulfur reservoir}
            
            The sulfur chemistry in the earliest phases of star formation, prestellar cores and very young Class 0 objects is particularly puzzling \citep{Fuente2016, Domenech2016}, where sulfur appears to be depleted several orders of magnitude from its cosmic abundance. The accretion of sulfur on the surface of dust grains was first considered as the cause of sulfur depletion. Since then, a lot of progress has been done in understanding the chemistry of sulfur over grain surfaces. Recent results \citep{Fuente2023} indeed point toward the S$^{+}$ ion to be a decisive actor in the sulfur depletion in molecular clouds. Once the sulfur is depleted onto grains, the identity of the sulfur reservoir in the solid phase is still elusive. H$_{2}$S is expected to be formed copiously by hydrogenation of sulfur in the ice matrix of grain surfaces. This possibility has been explored with the help of new chemical networks and models \citep[see, e.g.,][]{Vidal2017, Laas2019, NavarroAlmaida2020}. Our results show that this is indeed the case, with H$_{2}$S being the most abundant sulfur-bearing molecule. The solid H$_{2}$S may react with other compounds in the ice matrix. OCS is the second most abundant sulfur-bearing molecule. Its production may be enhanced even further as revealed by the recent laboratory work of \citet{elAkel2022}, where they predict high abundances of this molecule in the ice matrix produced by reactions that involve solid H$_{2}$S and solid CO. NS, H$_{2}$S$_{3}$, H$_{2}$CS, SO, and HCNS complete the list of the most abundant sulfur-bearing molecules. SO$_{2}$ gas-phase abundance in this component is $6.04\times 10^{-8}$, six times lower than that of SO. The abundance ratio SO$_{2}$/SO has been proposed as a chemical clock in protostellar environments \citep{Charnley1997, Hatchell1998}. The lower abundance of SO$_{2}$ compared to that of SO accounts for the young age of the protostellar object in this simulation.

            Our predictions on the identity of the most abundant elemental reservoirs is in agreement with the recent findings in interstellar ices that are being carried out with the James Webb Space Telescope \citep[JWST; see, e.g.,][]{McClure2023}. The detection of complex organic molecules is, however, still scarce, with methanol as the only confirmed detection so far. In the case of sulfur-bearing molecules, H$_{2}$S ice remains undetected as in previous estimations of its abundance \citep{JEscobar2011}, but this time with an upper limit of 0.6\% with respect to water. Unlike H$_{2}$S, OCS, one of our main sulfur reservoirs, is detected in interstellar ices \citep{Geballe1985, Palumbo1995}.
            
        \subsection{Chemical inheritance}\label{sec:chemicalInheritance}
        
            An important aspect to analyze about the chemical composition of the FHSC is to what extent the gas-phase chemical makeup of the hot component at the end of the simulation is a product of the collapsing core described by the simulation or highly dependent on the initial abundances set at the beginning of the simulation that are inherited from the prestellar cold core phase. The initial chemical abundances set at the beginning of the simulation were taken from the chemical modeling of the Barnard 1b core in \citet{NavarroAlmaida2020}. These initial gas and ice-phase abundances, as well as the gas-phase abundances of the hot component at the end of the simulation are shown in Fig. \ref{fig:twoPhaseModel}. As it is apparent from this plot, the ice-phase abundances of most of the elemental reservoirs at the beginning of the simulation are similar to the gas-phase abundances of the hot component predicted at the end of the MHD simulation. This suggests that the composition of the hot component is the result of thermal desorption of the initial ice content from the prestellar phase. There is, therefore, a strong chemical inheritance from the prestellar core phase to the hot core phase developed in later Class 0 objects. However, notable exceptions appear, including biologically important iCOMS such as methyl formate (HCOOCH$_{3}$), acetaldehyde (CH$_{3}$CHO), and formamide (NH$_{2}$CHO). Since these molecules were not formed abundantly in the prestellar core phase, characterized by high densities and cold temperatures, they are not expected to be formed during the cold core collapse described by the simulation. Instead, these molecules must be formed during the warm-up phase produced by the ignition of the protostar. These results are in agreement with the findings of \citet{Hincelin2013, Coutens2020}. The formation of these molecules is detailed in the following sections.
            
        \subsection{Magnetohydrodynamic simulations and two-step models}\label{sec:twoPhaseModel}

            Two-step models are commonly used in the literature to mimic the chemical evolution taking place in rapidly changing environments such as hot cores and corinos. As discussed before, the gas-phase abundances of the most abundant elemental reservoirs in the hot component of our MHD simulation indicate high chemical inheritance because the chemical makeup in this component is mostly the product of the thermal desorption of the initial ice content. However, molecules that are not inherited seem to be produced in the warm-up phase of the collapsing core. It is therefore reasonable to test if a simple two-step model comprising fixed prestellar core and hot dense core phases is able to reproduce the gas-phase abundances predicted by the more complex MHD simulation. In building this two-step model, we considered first the results of the chemical modeling of Barnard 1b in \citet{NavarroAlmaida2020} as the first phase in this model. The second phase was set as a hot phase, mimicking the warm-up occurring in the hot component at the end of the simulation. Since the distributions of gas-phase abundances in the hot component (Figs. \ref{fig:timeEvolutionAbs} and \ref{fig:histograms}) at the end of the simulation show little dispersion, we took the final physical conditions of a representative tracer particle belonging to the hot component as a template for the hot phase in this two-step model. As an upper bound, this phase spans the whole duration of the MHD simulation, $\sim 6.5\times 10^{4}$ yr, and has a fixed temperature of $500$ K and a constant density of $8\times 10^{-11}$ g cm$^{-3}$.

            In Fig. \ref{fig:twoPhaseModel} we show the comparison between the gas-phase abundances predicted by the two-step model and the resulting gas-phase abundances at the end of the MHD simulation. Among the carbon reservoirs, the greatest difference between the two-step model and the MHD simulation is found in CO, overproduced by the two-step model by a factor of two. Consequently, the two-step model underestimates the gas-phase abundances of other carbon elemental reservoirs such as formaldehyde ($\sim 0.3$ times less abundant), methanol ($\sim 0.5$ times lower), or methyl formate ($\sim 0.5$ times lower). In general, the carbon elemental abundances shown in Fig. \ref{fig:twoPhaseModel} are well reproduced by the two-step model. This is also the case for the oxygen elemental reservoirs. The largest difference between the gas-phase abundances of the MHD simulation and the two-step model among the oxygen elemental reservoirs is found in nitric oxide NO, with an underestimation from the two-step model by a factor of six. In the case of nitrogen and sulfur reservoirs, the two-step model performance is worse. The largest deviation from the MHD simulation predictions is found in nitrogen sulfide NS, with a factor of $\sim 20$ of difference. To summarize, the two-step model describes fairly well the gas-phase abundances of the carbon and oxygen reservoirs of Fig. \ref{fig:twoPhaseModel}. However, this description is not as accurate for the nitrogen and sulfur elemental reservoirs considered here.

        \subsection{Chemical production of the FHSC}\label{sec:chemProduction}

\begin{figure*}
                \centering
                \includegraphics[width =\textwidth]{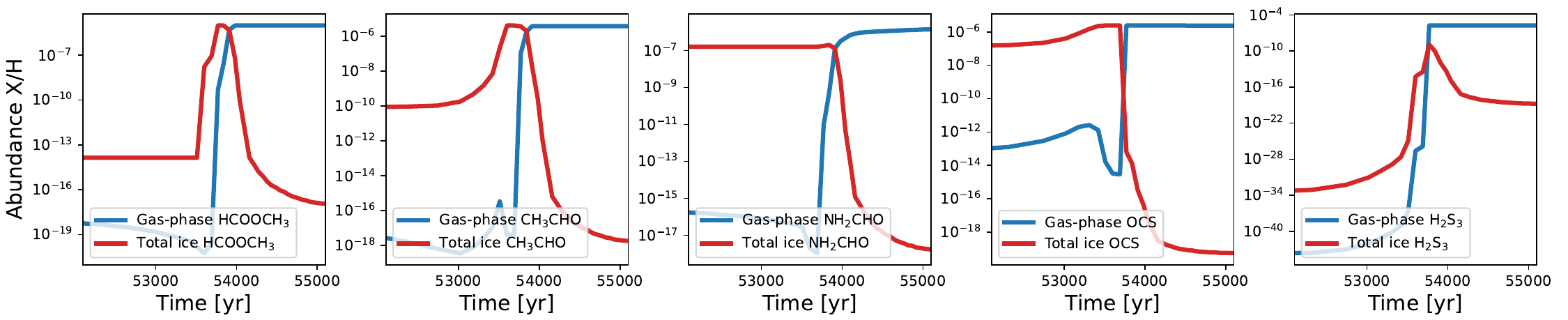}
                \caption{Time evolution of gas-phase (blue) and ice-phase (orange) chemical abundances of methyl formate (first column), acetaldehyde (second column), formamide (third column), OCS (fourth column), and H$_{2}$S$_{3}$ (fifth column) in the hot component.}
                \label{fig:methylFormate}
            \end{figure*}

            Despite the high degree of chemical inheritance, there are several molecules that were not present in the initial abundances, but they appear among the most abundant in the chemical composition of the hot component. The initial gas and ice-phase abundances of methyl formate (HCOOCH$_{3}$) are well below the gas-phase abundances predicted by the MHD collapse at the end of the simulation. Since the cold core chemistry of Barnard 1b, as modeled in \citet{NavarroAlmaida2020}, is not able to produce this molecule, methyl formate must be formed during the warm-up of the collapsing core. To check this, we examined the evolution of the gas-phase and ice-phase abundances of this molecule (see Fig. \ref{fig:methylFormate}). According to its evolution, this molecule is mostly present in ices before $\sim 5.3\times 10^{4}$ yr. Its ice-phase abundance is enhanced when the collapsing core increases its temperature. According to our chemical network, the total ice-phase abundance of this molecule rises as the rate of the diffusion reaction
            \begin{equation*}
                {\rm HCO}^{\rm ice} + {\rm CH}_{3}{\rm O}^{\rm ice} \rightarrow {\rm HCOOCH}_{3}^{\rm ice}
            \end{equation*}
            is enhanced by the rising temperatures of the collapse. Once the temperature is high enough, icy methyl formate is desorbed via thermal desorption. The peak of ice-phase methyl formate matches its final gas-phase abundance, meaning that no other mechanism is forming gas-phase methyl formate efficiently. The importance of this reaction to  form methyl formate in the warm-up phase of hot cores is discussed in \citet{Garrod2006}. Our estimation for the abundance of methyl formate is in agreement with measurements of this molecule in comets. Its abundance is found to be between $0.01-0.08$ relative to that of water \citep{Despois2005, Remijan2006} while our MHD simulation yields a ratio HCOOCH$_{3}$/H$_{2}$O $\sim 0.056$, well within the limits.

            A similar behavior is seen in acetaldehyde (CH$_{3}$CHO), with solid and gas-phase initial abundances well below the gas-phase abundances seen at the end of the simulation. The time evolution of its ice and gas-phase abundance reveals a similar pattern to that of methyl formate, with the molecule being mainly locked in ices before $\sim 5.3\times 10^{4}$ yr. Its ice-phase abundance is only increased when the gas and dust of the collapsing core start heating up. Our chemical network indicates that the ice-phase abundance of acetaldehyde is increased by the diffusion reaction
            \begin{equation*}
                {\rm HCO}^{\rm ice} + {\rm CH}_{3}^{\rm ice} \rightarrow {\rm CH}_{3}{\rm CHO}^{\rm ice},
            \end{equation*}
            whose rate grows with temperature. The icy acetaldehyde is completely thermally desorbed when temperatures reach several $\sim 100$ K. As with methyl formate, the peak in acetaldehyde ice abundance coincides with its final gas-phase abundance and therefore no other reactions are expected to form gas-phase acetaldehyde significantly besides the thermal desorption of CH$_{3}$CHO ice. The formation of this molecule in interstellar ices via combination of radicals has been thoroughly investigated in the literature, showing complex dependences on grain properties and temperature \citep{EnriqueRomero2021}. Our chemical network considers a diffusion over binding energy ratio $f=0.4$ that favors the reaction above to produce high abundances of this molecule. Apart from methyl formate and acetaldehyde, the gas-phase abundance of the remaining oxygen reservoirs at the end of our MHD simulation are comparable to the initial ice abundances, suggesting little production of these molecules during collapse.

            Among the most abundant nitrogen-bearing molecules, the initial ice abundances of formamide (NH$_{2}$CHO) and HCNO are below the gas-phase abundances seen in the MHD simulation. The dense core phase before the collapse is able to produce an ice abundance of formamide of $\sim 1.5\times 10^{-7}$ that is then increased up to $2.3\times 10^{-6}$. To analyze the formation of this molecule, we show the time evolution of its gas and ice-phase abundances in Fig. \ref{fig:methylFormate}. Unlike methyl formate or acetaldehyde, formamide ice-phase abundance does not increase to match the final gas-phase abundance when dust temperature is increased. This suggests that there should be other reactions responsible for the formation of gas-phase formamide besides the thermal desorption of its ice content. Our chemical network indicates that gas-phase reactions contribute to the production of gas-phase formamide at moderate temperatures $\sim 20$ K. In these conditions, the gas-phase reaction 
            \begin{equation*}
                {\rm NH_{2}}^{\rm gas} + {\rm H}_{2}{\rm CO}^{\rm gas} \rightarrow {\rm NH}_{2}{\rm CHO}^{\rm gas} + {\rm H}^{\rm gas}.
            \end{equation*}
            is the second most important path for the production of formamide. Combined with the thermal desorption of ices at higher temperatures, these two pathways account for most of the total gas-phase abundance predicted by the MHD simulation. As discussed in previous sections, the main gas-phase route of formation of formamide is in agreement with theoretical chemical calculations \citep{Barone2015}, and recent observational results support this relationship between formaldehyde and formamide \citep{Taniguchi2023}. Our chemical network does not include the recent findings of formamide formation at low temperatures in H$_{2}$O and CO-rich ices under cold molecular cloud conditions presented in \citet{Chuang2022}. HCNO is another nitrogen-bearing molecule whose initial ice abundance is much lower than the gas-phase abundance seen at the end of the simulation. According to our chemical network, the thermal desorption of HCNO ice and the gas-phase neutral-neutral reaction of methylene (CH$_{2}$) with nitric oxide, a process already proposed for the formation of HCNO \citep{Marcelino2009}, are responsible for the final gas-phase abundance predicted by our simulation.

            The gas-phase abundance of OCS found in our simulation is higher than the initial ice content. This implies additional formation routes besides the thermal desorption of the initial ice content. The time evolution of gas and ice-phase abundances shows an increase in the ice abundance of OCS right before the protostar reaches temperatures of a few $\sim 100$ K, in line with the behavior of methyl formate or acetaldehyde (Fig. \ref{fig:methylFormate}). In this case, the diffusion reactions
            \begin{equation*}
                {\rm S}^{\rm ice} + {\rm CO}^{\rm ice} \rightarrow {\rm OCS}^{\rm ice}
            \end{equation*}
            and
            \begin{equation*} 
                {\rm O}^{\rm ice} + {\rm CS}^{\rm ice} \rightarrow {\rm OCS}^{\rm ice}
            \end{equation*}
            are responsible for the increase of the total ice-phase abundance of OCS. When temperatures of a few $\sim 100$ K are reached, the solid OCS is thermally desorbed. Again, since the ice abundance peak matches the final gas-phase abundance, there must not be additional mechanisms producing gas-phase OCS significantly. The gas-phase abundance of H$_{2}$S$_{3}$ predicted by the MHD simulation is, according to our model, the result of the thermal desorption of the high H$_{2}$S$_{3}$ ice abundances formed by thermal diffusion reactions occurring over grain surfaces:
            \begin{equation*}
                {\rm HSSH}^{\rm ice} + {\rm HS}^{\rm ice} \rightarrow {{\rm H}_{2}{\rm S}_{3}}^{\rm ice} + {\rm H}^{\rm ice}
            \end{equation*}
            and
            \begin{equation*} 
                {\rm HSS}^{\rm ice} + {\rm HS}^{\rm ice} \rightarrow {{\rm H}_{2}{\rm S}_{3}}^{\rm ice}.
            \end{equation*}
            However, our chemical network only includes thermal desorption of H$_{2}$S$_{3}$ ice as the formation route for gas-phase H$_{2}$S$_{3}$. Therefore, the fact that the H$_{2}$S$_{3}$ ice-phase abundance peak is below the final gas-phase abundance means that once the H$_{2}$S$_{3}$ ice is formed, following the reactions above, it is immediately desorbed from the grain surface, not contributing to an enhancement of the total ice abundance of this molecule. Finally, the gas-phase abundance of HCNS at the end of the MHD simulation is much higher than the initial gas and ice-phase abundances. As with HCNO, this molecule is mainly produced via gas-phase reactions involving methylene and, in this case, NS.

        \section{Grain growth and its chemical impact}\label{sec:grainGrowth}

            As it is apparent from the results obtained in Sect. \ref{sec:discussion}, grain surface reactions have a critical importance for the chemical composition in the FHSC. In spite of the high dependence on the initial chemical abundances, highly abundant complex organic molecules are formed via grain-phase reactions during the warm-up of the FHSC described by our MHD simulation. It is therefore worth investigating whether the chemical abundances we predicted here change as dust properties vary. In our analysis using the astrochemical code \texttt{NAUTILUS}, dust grains were regarded as uniform silicate spheres with a fixed average radius $\bar{r} = 0.1\ \mu$m, a mass density of $\rho = 3$ g cm$^{-3}$, a total dust-to-gas mass ratio of $\overline{R_{DG}} = 10^{-2}$, and a site density of $10^{15}$ cm$^{-2}$. However, some of these assumptions do not realistically describe the nature of dust during the formation of stars. Recent observations suggest that dust grows in a significant manner even before the formation of a protoplanetary disk \citep[see, e.g.,][]{Pagani2010, Galametz2019}.

            \subsection{Physical history of dust}\label{sec:grainGrowthPhysical}
        
                Aiming to explore how dust grains grow during the formation of a FHSC and how that affects its chemical makeup, we used the code \texttt{SHARK} \citep{Lebreuilly2023}. We introduced a module in \texttt{SHARK} in such a way that, instead of computing the changes in the dust grain size distribution and solving the hydrodynamical equations for the mixture of gas and dust self-consistently, grain sizes evolve following the time evolution of density, temperature, and magnetic field given by the physical history of the tracer particles from our MHD simulation. Even though \texttt{RAMSES} and \texttt{SHARK} work with different degrees of freedom, the information given to \texttt{SHARK} is scalar, and therefore can be introduced in the solvers for dust growth and fragmentation in 1D. Since the dynamics of the collapse is already fixed to what it was obtained in the simulation of \citet{Hennebelle2016}, dust properties do not affect it. The treatment of turbulence as a mechanism producing dust differential velocities follows the low-mass dense core collapse models of \citet{Guillet2020} and \citet{Ormel2009}, in which the injection scale of turbulence is equal to the Jean's mass, with an injection velocity equal to the sound speed. Turbulence is transferred to smaller scales following the Kolmogorov cascade of turbulence, which is controlled by the Reynolds number and depends on the density and temperature:
                \begin{equation}
                    {\rm Re} = 6.2\times 10^{7}\sqrt{\frac{n_{\rm H}}{10^{5}\ {\rm cm}^{-3}}}\sqrt{\frac{T}{10\ {\rm K}}}.
                \end{equation}
                Because this model describes the role of turbulence as a trigger for dust differential velocities in low-mass cores, like the one we study here, this prescription may not be appropriate for massive star formation.
                
\begin{figure*}
                    \centering
                    \includegraphics[width =\textwidth]{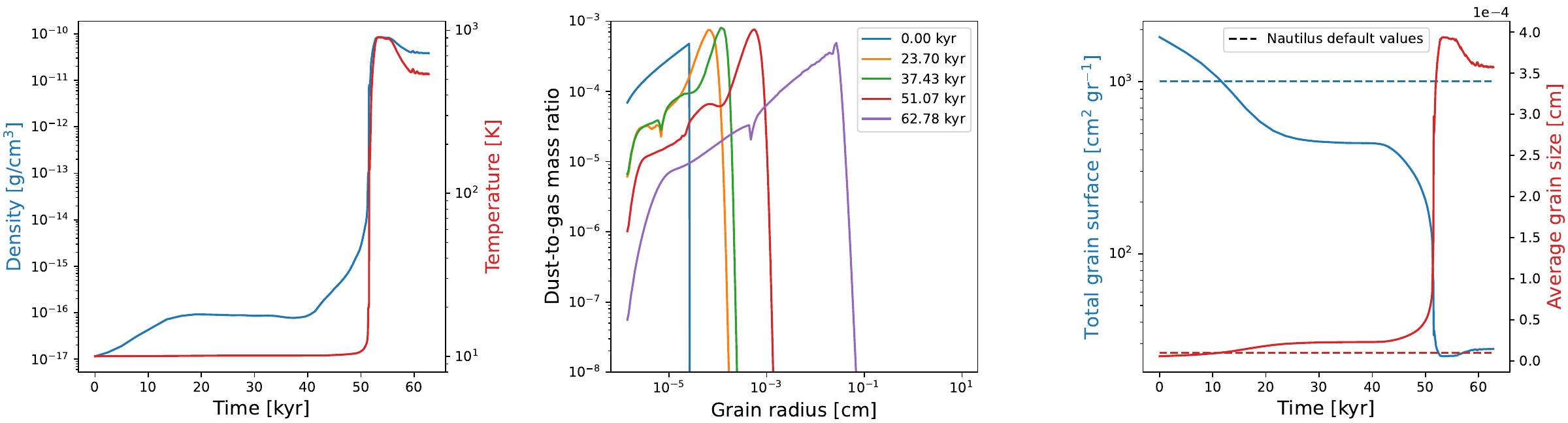}
                    \caption{Time evolution of physical properties and their impact on grain size in the hot component. \emph{Left:} Evolution of density (blue) and temperature (red) of the tracer particle selected to represent the hot component. \emph{Middle:} Dust size distribution at several snapshots. \emph{Right:} Time evolution of the total adsorption surface (blue solid line) and average grain size (red solid line). Also shown are the default \texttt{NAUTILUS} values for the total surface (blue dashed line) and average grain size (red dashed line) used in the chemical analysis in Sects. 4 and 5.}
                    \label{fig:grainGrowthHot}
                \end{figure*}

                With the physical information provided by tracer particles, we computed the time evolution of the grain-size distribution. The initial grain-size distribution follows the MRN distribution \citep[Mathis, Rumplr, and Nordsieck;][]{Mathis1977}, with grain sizes $r$ following the power law $\sim r^{-3.5}$ and ranging between $r_{\rm min}=0.05\ \mu$m and $r_{\rm max}=10$ cm. The dust-to-gas mass ratio was set to $10^{-2}$ and the grain material density to 3 g cm$^{-3}$. Now we are going to examine the grain growth taking place in the temperature components and scales we defined in Sect. \ref{sec:results}. In the left panel of Fig. \ref{fig:grainGrowthHot} we show the time evolution of density and temperature of the tracer particle belonging to the hot component that was selected as representative in the analysis presented in Sect. \ref{sec:discussion}. After processing the physical evolution of the tracer particle with \texttt{SHARK}, we obtained the dust size distribution at several time steps in the middle panel of Fig. \ref{fig:grainGrowthHot}. Dust grains do not grow significantly beyond 0.1 $\mu$m during the first $\sim 30$ kyr, when the grain size distribution peaks at $\sim 2\times 10^{-4}$ cm, that is, an order of magnitude higher compared to the initial value. It is only with densities higher than $10^{-14}$ g cm$^{-3}$, at $\sim 50$ kyr, when dust sizes start increasing rapidly. This is consistent with the results presented in \citet{Navarro2023}, where the analysis of the spectral index, a proxy of grain size, concluded that grain growth would only be present in evolved Class 0 and I YSOs. In the last $\sim 10$ kyr, once gas density reaches $\sim 10^{-11}$ g cm$^{-3}$, the grain-size distribution peak occurs at a grain radius of $\sim 2\times 10^{-2}$ cm, three orders of magnitude higher than the initial value. This indicates that gas density, in its role on turbulence as the dominant mechanism responsible for grain growth \citep{Lebreuilly2023}, is an important driver of grain coagulation, in agreement with the findings of \citet{Ossenkopf1993} and \citet{Ormel2009}.

                Changes in dust grain sizes have an impact in grain thermodynamics and in the extent of adsorbing surface. As investigated by \citet{Iqbal2018}, the size-dependent surface temperature plays a key role in determining the chemical abundances of several molecules. In our case, however, dust grains are expected to be in thermal equilibrium with the gas due to the high densities of our environment. Therefore, we investigated the loss of adsorbing surface as a consequence of grain coagulation under the assumption of dust mass conservation. In such scenario, we computed the total surface of adsorption $S_{T}$ per gram of gas at each time step as
                \begin{equation}\label{eq:surface}
                    S_{T} = \frac{3}{\rho}\sum_{i}^{N} \frac{R_{DG}(r_{i})}{r_{i}},
                \end{equation}
                where $\rho$ is the dust material density, $R_{DG}(r_{i})$ is the dust-to-gas mass ratio for each grain radius $r_{i}$, and $N$ is the number of dust bins set in the calculations. In this case, $N=200$. The time evolution of adsorbing surface per gram of gas is shown in the right panel of Fig. \ref{fig:grainGrowthHot}. This confirms the decreasing total grain surface as a result of grain coagulation from $\sim 2\times 10^{3}$ cm$^{2}$ gr$^{-1}$ at the beginning of the simulation to $\sim 30$ cm$^{2}$ gr$^{-1}$ at the end. Given a total dust grain surface $S_{T}$ at a certain time step, we computed its corresponding average dust grain size $\bar{r}$ by assuming $r_{i} = \bar{r}$ for all values of $i$. Introducing this assumption into Eq. \ref{eq:surface} and taking into account that the sum of the dust-to-gas mass ratio of all bins yields the total initial dust-to-gas mass ratio $\overline{R_{DG}} = 10^{-2}$, we have:
                \begin{equation}\label{eq:averageRadius}
                    \bar{r} = \frac{3}{\rho}\frac{\overline{R_{DG}}}{S_{T}}.
                \end{equation}
                The time evolution of the average grain size $\bar{r}$ is also shown in Fig. \ref{fig:grainGrowthHot}, increasing up to $\sim 4\times 10^{-4}$ cm. So far, in the chemical analysis of the FHSC with the chemical code \texttt{NAUTILUS} (Sects. \ref{sec:results} and \ref{sec:reservoirs}-\ref{sec:chemProduction}), an average grain radius of $\bar{r}=10^{-5}$ cm and a constant dust-to-gas mass ratio of $\overline{R_{DG}} = 10^{-2}$ were provided. These parameters are the only supported inputs that account for dust size changes in the version of \texttt{NAUTILUS} we used. These default values are also shown in the right panel of Fig. \ref{fig:grainGrowthHot}. In this figure, the evolution of the total grain surface and average grain size computed with \texttt{SHARK} clearly depart from the default \texttt{NAUTILUS} values.

\begin{figure*}
                    \centering
                    \includegraphics[width =\textwidth]{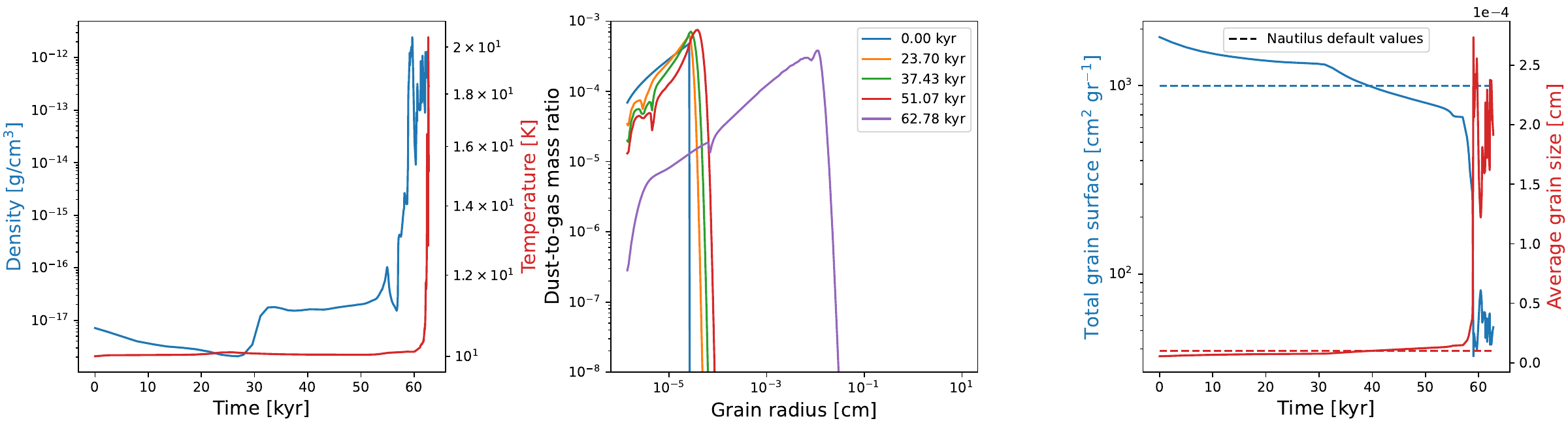}
                    \caption{Time evolution of physical properties and their impact on grain size in the cool component. \emph{Left:} Evolution of density (blue) and temperature (red) of the tracer particle selected to represent the cool component. \emph{Middle:} Dust size distribution at several snapshots. \emph{Right:} Time evolution of the total adsorption surface (blue solid line) and average grain size (red solid line). Also shown are the default \texttt{NAUTILUS} values for the total surface (blue dashed line) and average grain size (red dashed line) used in the chemical analysis in Sects. 4 and 5.}
                    \label{fig:grainGrowthWarm}
                \end{figure*}

                We performed a similar analysis for a representative particle of the cool component. We selected a tracer particle whose physical properties are shown in Fig. \ref{fig:grainGrowthWarm} and include both cold temperatures $<20$ K and densities up to $10^{-12}$ g cm$^{-3}$. This tracer particle is representative of the cool component as the ranges of density and temperature that it covers are similar than the median values shown in Fig. \ref{fig:timeEvolutionAbs}. After processing the physical history of this tracer particle with \texttt{SHARK}, we obtained the dust size distribution at several time steps (see Fig. \ref{fig:grainGrowthWarm}). In this case, due to the lower densities traced by this particle during most of its time evolution compared to the hot component, the grain size distribution does not change significantly until 50 kyr. The peak of the grain size distribution at times $t<50$ kyr is found around $\sim 3\times 10^{-5}$ cm and it does not vary significantly. At $t>50$ kyr, when the gas density rises (left panel of Fig. \ref{fig:grainGrowthWarm}), the peak of the grain size distribution is around $10^{-2}$ cm, three orders of magnitude higher than the initial size $10^{-5}$ cm. Applying Eqs. \ref{eq:surface} and \ref{eq:averageRadius}, we computed the time evolution of the total adsorbing surface and the average grain size (see Fig. \ref{fig:grainGrowthWarm}). As this particle traces lower densities throughout its history, it exhibits a moderate grain growth. The average grain size is increased in up to a factor of $\sim 20$ when densities reach $10^{-12}$ g cm$^{-3}$ (see Fig. \ref{fig:grainGrowthWarm}). This growth is a factor of two lower than that observed for the tracer particle belonging to the hot component.
            
\begin{figure*}
                    \centering
                    \includegraphics[width =\textwidth]{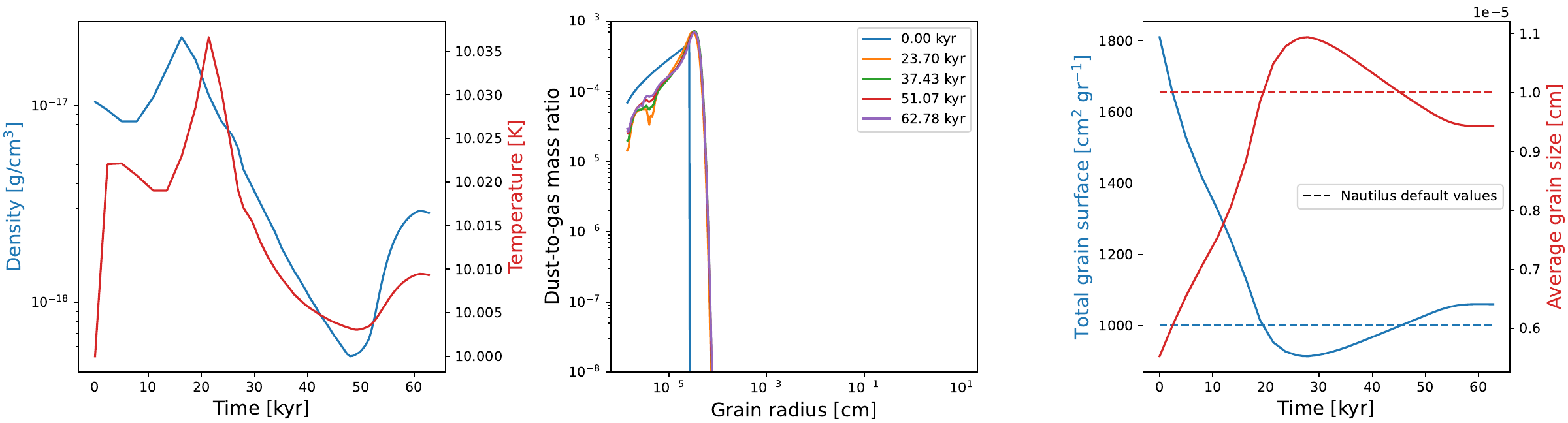}
                    \caption{Time evolution of physical properties and their impact on grain size in the cold component. \emph{Left:} Evolution of density (blue) and temperature (red) of the tracer particle selected to represent the cold component. \emph{Middle:} Dust size distribution at several snapshots. \emph{Right:} Time evolution of the total adsorption surface (blue solid line) and average grain size (red solid line). Also shown are the default \texttt{NAUTILUS} values for the total surface (blue dashed line) and average grain size (red dashed line) used in the chemical analysis in Sects. 4 and 5.}
                    \label{fig:grainGrowthCold}
                \end{figure*}

                Finally, we consider the case of a tracer particle that probes the temperature and density of the outermost parts of the FHSC, that is, a tracer particle representative of the cold component. The evolution of its physical properties, the grain-size distribution, the total grain surface, and the average grain size are given in Fig. \ref{fig:grainGrowthCold}. Unlike the tracer particles chosen to sample the cool and hot components, in this case there is no sudden increase of density and temperature near the end of the simulation. The time evolution is essentially isothermal with variations in density of up to two orders of magnitude. As a consequence, the grain size distribution only shows a five-fold increment in the maximum grain size. Moreover, this increase is reached early on, with no changes in the grain size distribution beyond $\sim 24$ kyr. Beyond $\sim 20$ kyr, the total dust grain surface and the average grain size are in good agreement with \texttt{NAUTILUS} default values for the total dust-to-gas mass ratio $\overline{R_{DG}} = 10^{-2}$ and average grain size $\bar{r}=10^{-5}$ cm we used in the chemical analysis of Sects. \ref{sec:results} and \ref{sec:discussion} (see Fig. \ref{fig:grainGrowthCold}). These default values are therefore appropriate to describe the dust and chemistry of the envelope of the FHSC.

            \subsection{Impact of grain growth on molecular abundances}\label{sec:chemGrainGrowth}
            
\begin{figure*}
                    \centering
                    \includegraphics[width =\textwidth]{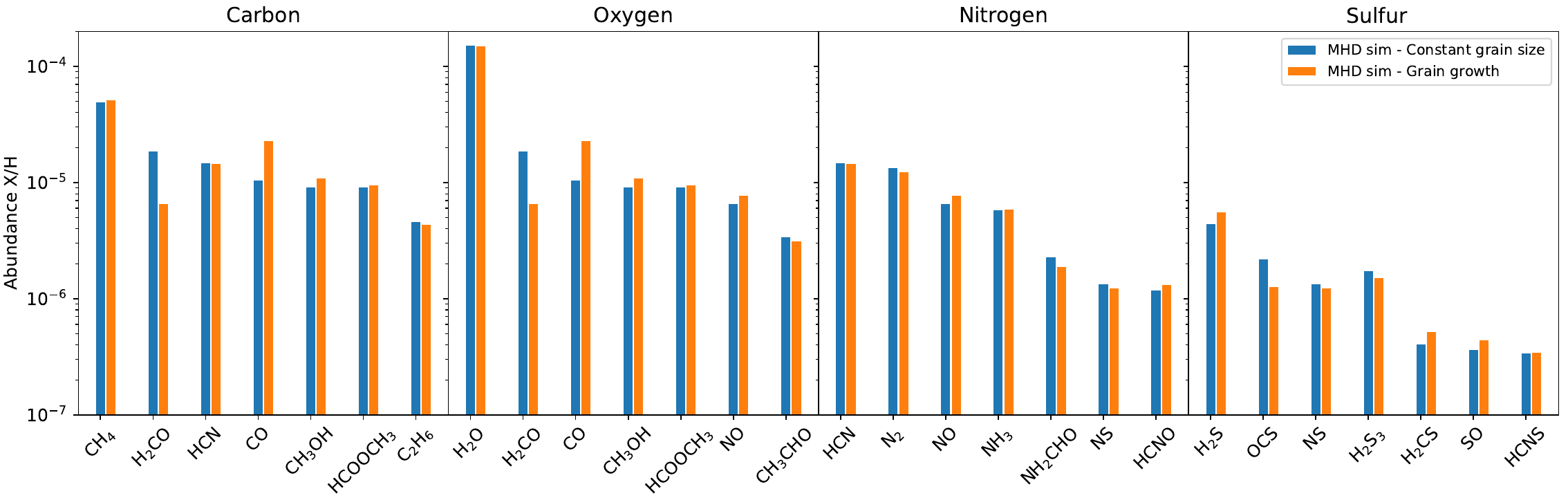}
                    \caption{Comparison between the gas-phase abundances of the most abundant elemental reservoirs produced at the hot core by the MHD simulation (green bars) and those obtained introducing the evolution of the average grain size shown in Fig. \ref{fig:grainGrowthHot}.}
                    \label{fig:histogramGrainGrowth}
                \end{figure*}

                The input parameters that control dust grain size in the gas-grain chemical code \texttt{NAUTILUS} are the dust-to-gas mass ratio and the grain radius. The version of \texttt{NAUTILUS} we used does not allow the treatment of grain size distributions. However, eqs. \ref{eq:surface} and \ref{eq:averageRadius} allowed us to incorporate grain size changes in the chemical model. First, for a given tracer particle, \texttt{SHARK} is used to obtain the time evolution of the dust-size distribution. At any given time step, the dust-size distribution is introduced into Eq. \ref{eq:surface} to provide the total adsorption surface. The total adsorption surface at an instant is then inserted into Eq. \ref{eq:averageRadius}. In this equation, the average grain size $\bar{r}$ and the total dust-to-gas mass ratio $\overline{R_{DG}}$ can be chosen freely so that they yield the total adsorption surface calculated before. In the following, we assumed $\overline{R_{DG}} = 10^{-2}$ and let the average grain size $\bar{r}$ vary so that we get the total dust grain surface dictated by \texttt{SHARK}. Therefore, adjusting the dust grain size during the computation of abundances with \texttt{NAUTILUS}, we were able to include grain growth in the chemical code. As a first comparison, in Fig. \ref{fig:histogramGrainGrowth} we compare the gas-phase abundances of the main elemental reservoirs in the hot component as calculated previously on Fig. \ref{fig:twoPhaseModel}) with the same abundances once grain growth is included in \texttt{NAUTILUS}. Given that the prevalent chemical processes in the core of the FHSC are thermal diffusion and thermal desorption, it is not surprising that the loss of adsorbing/desorbing surface has a little impact in the abundances of the main elemental reservoirs. The results show that the largest differences are in the abundances of H$_{2}$CO, with a difference of a factor of $\sim 3$; CO, a factor of $\sim 2$; and OCS, a factor of $\sim 1.7$ (see Fig. \ref{fig:histogramGrainGrowth}). This suggests that grain growth may hinder the formation of H$_{2}$CO by accretion of gas-phase CO right before being thermally desorbed.

\begin{figure*}
                    \centering
                    \includegraphics[width =\textwidth]{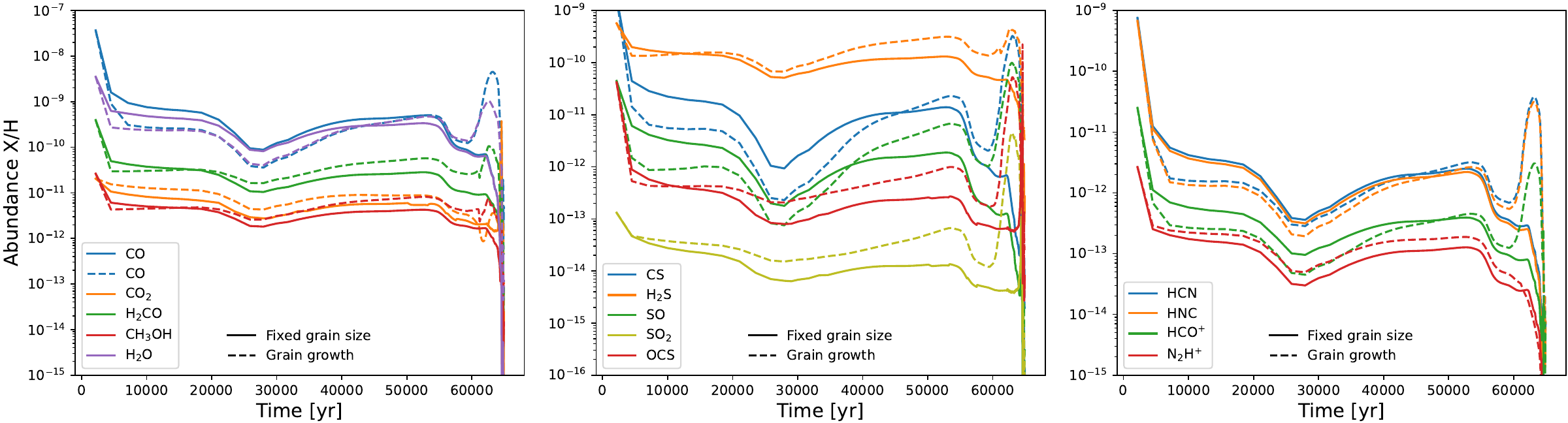}
                    \caption{Time evolution of gas-phase abundances of several molecules as probed by the selected particle of the cool component considering a fixed grain size (solid lines) and including the grain growth obtained in Fig. \ref{fig:grainGrowthWarm} (dashed lines).}
                    \label{fig:timeEvGrainGrowth}
                \end{figure*}

                Molecular depletion is a prevalent chemical process for tracer particles in dense and cold environments. This process is expected to be affected by grain growth and the subsequent loss of adsorbing surface, leading to higher gas-phase abundances. As we saw in Sect. \ref{sec:results}, the chemistry of the cool component is heavily influenced by depletion. To investigate the effects of grain growth in the chemistry of this component, we computed the chemical evolution of the gas-phase abundances of the molecules presented in Fig. \ref{fig:timeEvolutionAbs} taking into account the increasing average grain size as dictated by the tracer particle representative of the cool component (Sect. \ref{sec:grainGrowthPhysical}). In Fig. \ref{fig:timeEvGrainGrowth} we compare the gas-phase abundances computed with fixed grain size (Sect. \ref{sec:results}) and the ones resulting from the inclusion of grain growth in the cool component. After $\sim 60$ kyr, when grain growth becomes significant (see Fig. \ref{fig:grainGrowthWarm}), there is a clear difference between the gas-phase abundances estimated using a fixed grain size and the one estimated with varying grain sizes. In general, grain growth leads to a delay in the depletion of gas-phase abundances and a great enhancement of gas-phase abundances for most molecules. This enhancement reaches up to a factor of $\sim 100$ for molecules such as CO, CS, OCS, SO, SO$_{2}$, HCN, and HNC. The enhancement is less prominent in the case of formaldehyde (H$_{2}$CO), methanol (CH$_{3}$OH), H$_{2}$S, N$_{2}$H$^{+}$, and CO$_{2}$. The behavior of the gas-phase abundances of the sulfur-bearing molecules SO, SO$_{2}$, and H$_{2}$S is in agreement with the results obtained by \citet{Harada2017}. With the loss of adsorbing surface, molecules formed by hydrogenation and chemical desorption like H$_{2}$S are produced less efficiently, leading to a much lower increase in the gas-phase abundance compared to that of SO or SO$_{2}$. Although significant, the enhancement of gas-phase abundances produced by grain growth is short lived $\sim 4$ kyr and abundances become depleted again due to the increasing density (Fig. \ref{fig:grainGrowthWarm}).
                
                Finally, as we saw in Sect. \ref{sec:grainGrowthPhysical}, for a tracer particle representative of the cold component, the values of the total adsorption surface and average grain size do not depart significantly from the default values used in the chemical analysis of Sect. \ref{sec:results}. Consequently, the results presented for the cold component in Sect. \ref{sec:results} offer a fair description of the chemistry in the outermost areas of the FHSC.

    \section{Summary and conclusions}

        In this paper we have presented the chemical modeling of a state-of-the-art non-ideal 3D MHD simulation of a solar-mass collapsing core progressing up to the formation of a FHSC. The tracer particles included in the simulation allowed us to probe the evolution of physical properties throughout the collapse. These tracer particles were divided into different sets according to their temperature at the end of simulation to describe the relevant chemical processes occurring on several spatial scales. We postprocessed the particle histories with a gas-grain chemical code to investigate the chemical composition of the FHSC. Additionally, we included grain growth in our postprocessing to examine the impact of this phenomenon in the chemical makeup of the collapsing core. In more detail:
        
        \begin{itemize}
        
            \item We used a non-ideal MHD simulation of a solar-mass collapsing core during 64 kyr to study the chemical evolution during the core collapse to form a FHSC. The initial setting of our simulation is a 1 $M_{\odot}$ strongly magnetized (initial mass-to-flux ratio of $\mu=2$) and turbulent (velocity fluctuations match an initial Mach number $\mathcal{M}=1.2$) sphere of radius $r\sim 2500$ au, uniform density of $\rho= 10^{-17}$ g cm$^{-3}$, and uniform temperature of 10 K.
            
            \item We analyzed the evolution of chemical abundances in our simulation using tracer particles. In particular, we followed the time evolution of $2\times 10^{5}$ particles that sampled the physical conditions of the protostellar envelope and compact disk components. However, the outflow was poorly sampled and its chemistry is not studied in this work. The particles were postprocessed with the gas-grain chemical code \texttt{NAUTILUS}, setting the chemical composition of the Barnard 1b core as initial abundances. To analyze the different chemical processes and scales, we divided the tracer particles into different regions according to their temperature at the end of the simulation.
            
            \item We examined the chemical composition of the hot core at the end of the simulation looking at the most abundant elemental reservoirs of carbon, oxygen, nitrogen, and sulfur. We studied the chemical inheritance by comparing the gas-phase hot core abundances with the initial gas and ice abundances set at the beginning of the simulation. Our results indicate that there is a strong chemical inheritance. However, some iCOMs of biological interest such as methyl formate, acetaldehyde, and formamide are only formed abundantly during the warm-up phase of the FHSC by gas-phase reactions in hot gas and the enhancement of diffusion reactions occurring in the surface of dust grains.
            
            \item Since there is a strong chemical inheritance, that is, the chemical abundances of the hot core at the end of the simulation are mostly the product of the thermal desorption of the initial icy abundances, we tested whether we could replicate the chemical abundances seen at the end of the simulation with a simple two-step model (cold+warm). Our results show that this two-step model reproduces the chemical abundances of the most abundant carbon and oxygen elemental reservoirs reasonably well, with differences up to a factor of two, but it does not for nitrogen and sulfur elemental reservoirs, with differences of up to a factor of $\sim 20$. This implies that these molecules are more sensitive to the dynamical evolution described by the simulation.

            \item For the first time, we investigated the role of the grain size in the chemical makeup of the collapsing core. We used the code \texttt{SHARK} and the physical history probed by the tracer particles to obtain the time evolution of the dust size distribution in different areas of the FHSC. In low density areas, grains do not grow significantly, with average sizes around $10^{-5}$ cm, similar to the size used in our chemical analysis. At distances of $\sim 40$ au from the center, we observed that the average grain size is increased by a factor of $\sim 20$. In the hot core, grain growth proceeds rapidly to reach a maximum grain size of $\sim 0.1$ cm in 64 kyr.
            
            \item We used the chemical code \texttt{NAUTILUS} to investigate the impact of grain growth on the chemical composition of the protostar. The major effect of grain  growth is observed in the cool and dense component where the temperature is below $\sim 100$ K and adsorption on grain surfaces is still driving the chemistry. At distances of $\sim 40$ au from the center, the loss of adsorption surface produced by grain growth is translated in an enhancement of gas-phase abundances, as this loss is able to delay the depletion of molecules. This enhancement varies among the molecules analyzed in this paper, and therefore molecular abundance ratios can be used as grain growth proxies. The chemical composition of the hot core is not substantially affected by grain growth; the greatest differences are found in the abundances of H$_{2}$CO, decreased by around a factor of four, and CO, increased by a factor of two.
  
        \end{itemize}
        
        With the aid of numerical simulations, we investigated the chemistry of the FHSC phase with emphasis on chemical inheritance and production, the suitability of simpler models commonly used in the literature to describe their chemistry, and the role of grain growth in their composition. Synthetic maps of the molecular emission of the FHSC and dust opacity maps taking into account grain growth are the next natural steps toward a better comprehension of the observational features these objects exhibit. This effort would be an invaluable tool for observers to detect this elusive stage in the star formation process.
        
    \begin{acknowledgements}
        DNA acknowledges funding support from Fundaci\'on Ram\'on Areces through its international postdoc grant program. DNA also acknowledges the computer resources and assistance provided by Centro de Computación Científica-Universidad Autónoma de Madrid (CCC-UAM). PH and UL acknowledge funding from the European Research Council synergy grant ECOGAL (Grant : 855130). DNA, AF, and PRM thank the Spanish MICIN for funding support from PID2019-106235GB-I00. AF acknowledges funding from the European Research Council advanced grant SUL4LIFE (GAP-101096293). BC acknowledges funding from the French national agency ANR DISKBUILD (ANR-20-CE49-0006). VW acknowledges the CNRS program "Physique et Chimie du Milieu Interstellaire" (PCMI) co-funded by the Centre National d'Etudes Spatiales (CNES). YA acknowledges financial support from grant PID2019-107408GB-C42 of the Spanish State Research Agency (AEI/10.13039/501100011033). 
    \end{acknowledgements}

    \bibliography{FHSC_I.bib}
    \bibliographystyle{aa}

\begin{appendix}
\onecolumn
    \section{Gas-phase abundance distributions}\label{sec:appendixHistograms}
    In this appendix we plot the statistical distribution of gas-phase abundances at the end of the simulation for the temperature components defined in Sect. \ref{sec:tempComponents}, their first and third quartiles, and the median.
        \begin{figure}[h]
            \centering
            \includegraphics[width =\textwidth]{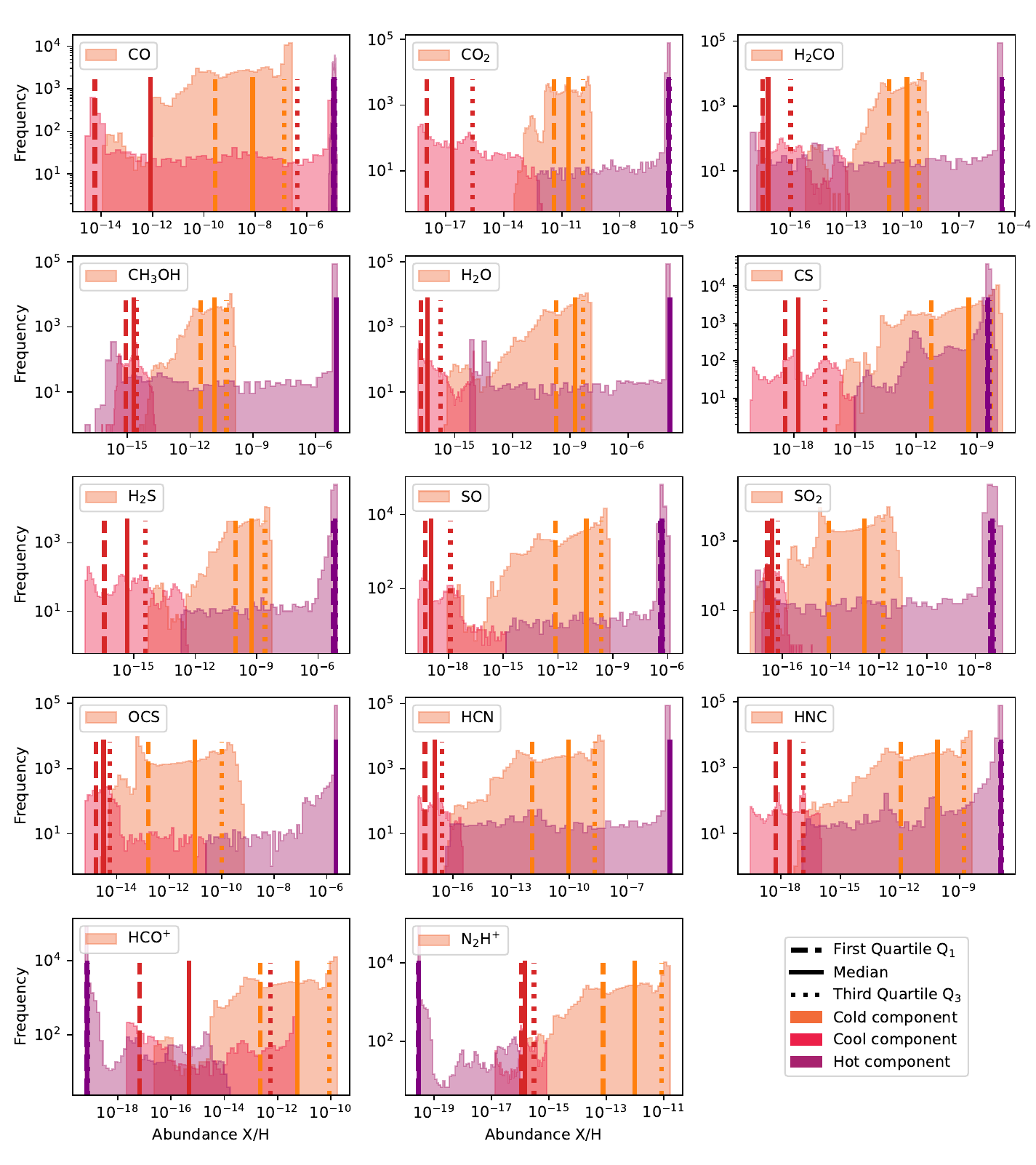}
            \caption{Statistical distributions of gas-phase abundances for each temperature component. The first and third quartiles are shown, as well as the median. The distribution for each temperature component has 50 bins.}
            \label{fig:histograms}
        \end{figure}
    \newpage
    \section{Spatial distribution of chemical abundances}\label{sec:spatialDist}
        In this section we plot the spatial distribution of the tracer particles and their gas-phase and ice-phase abundances of the molecules that appear in Fig. \ref{fig:timeEvolutionAbs}.
        \subsection{CO\texorpdfstring{$\mathsf{_{2}}$}\ \ and H\texorpdfstring{$\mathsf{_{2}}$}\ CO}
        \vfill
        \begin{figure}[h]
            \centering
            \begin{subfigure}{\textwidth}
                \centering
                \includegraphics[width=\textwidth]{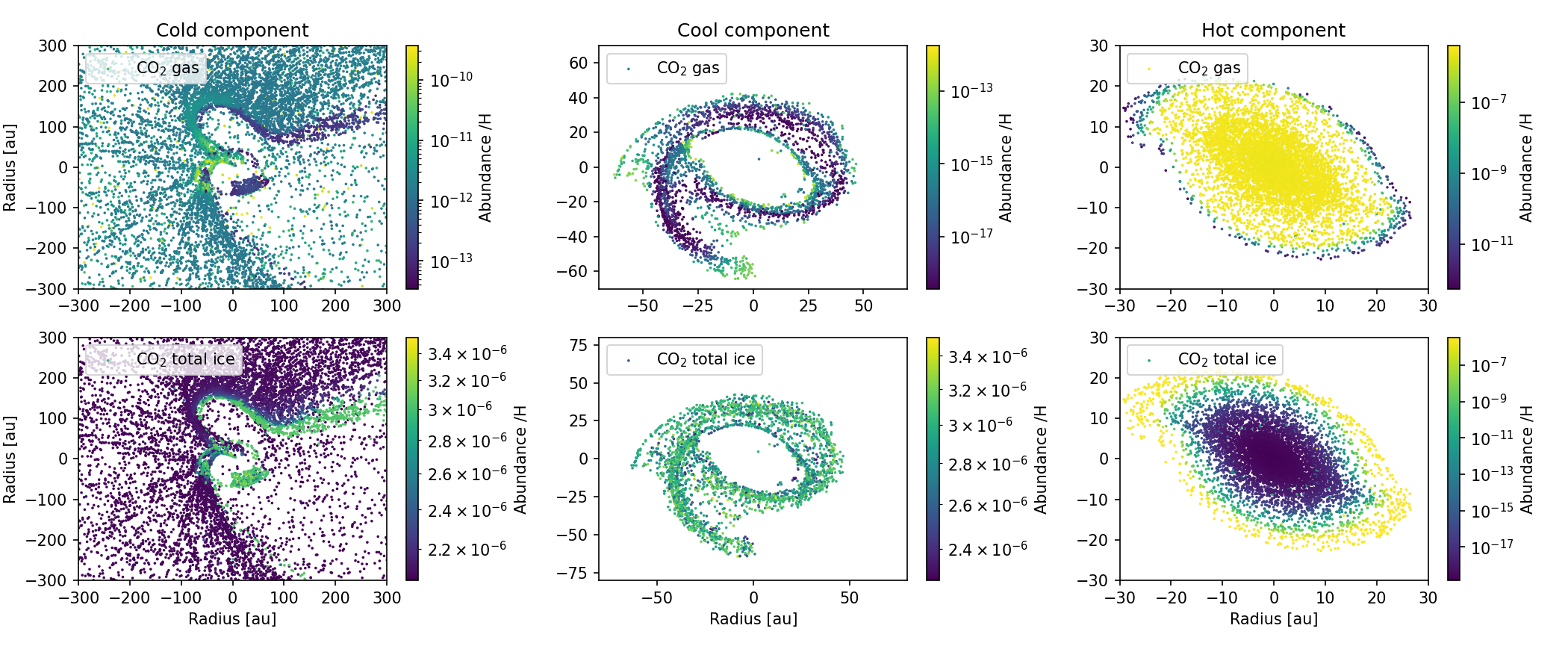}
                \caption{Spatial distribution of the tracer particles at the end of the simulation and their CO$_{2}$ gas-phase (top row) and total ice-phase (bottom row) abundances  for the three temperature components: cold (left), cool (middle), and hot (right).}
            \end{subfigure}
            $\left.\right.$\\
            $\left.\right.$\\
            $\left.\right.$\\
            $\left.\right.$\\
            $\left.\right.$
            \begin{subfigure}{\textwidth}
                \centering
                \includegraphics[width=\textwidth]{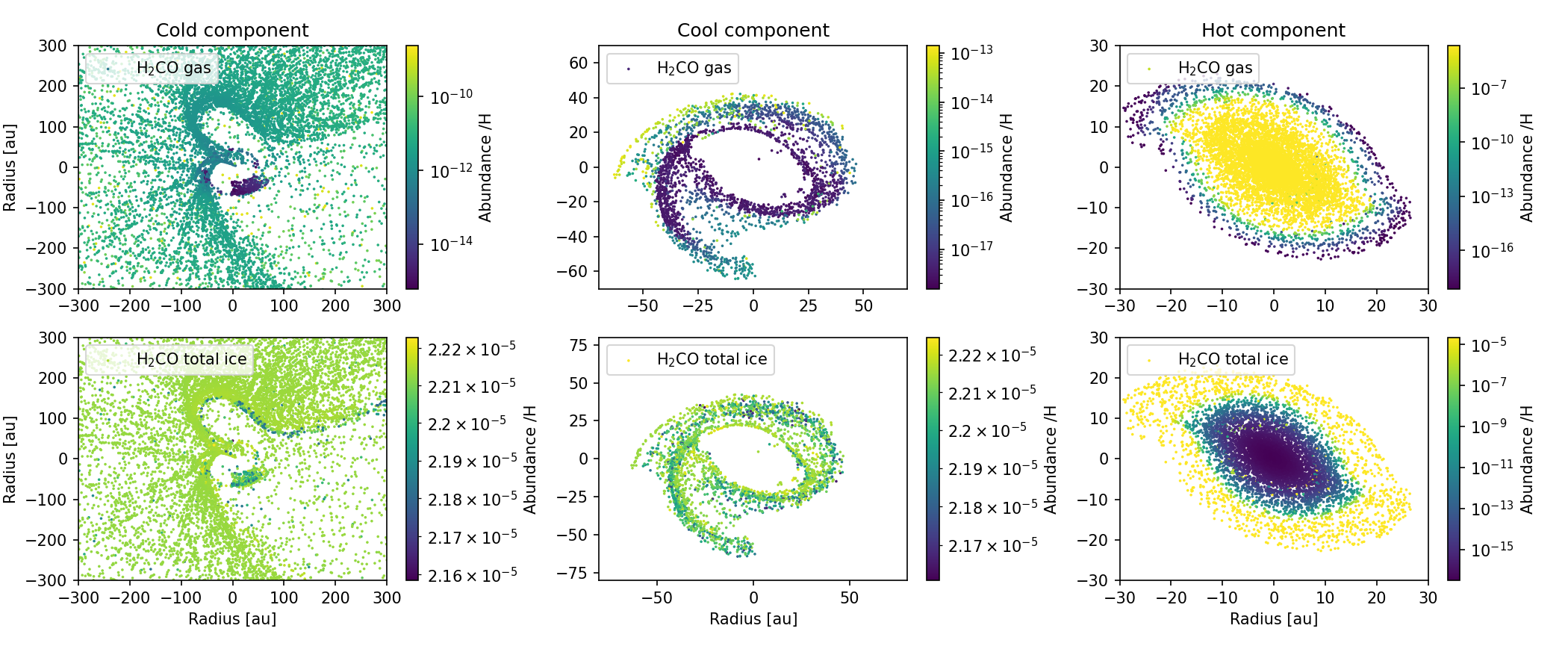}
                \caption{Spatial distribution of the tracer particles at the end of the simulation and their H$_{2}$CO gas-phase (top row) and total ice-phase (bottom row) abundances  for the three temperature components: cold (left), cool (middle), and hot (right).}
            \end{subfigure}
        \end{figure}
        \vfill
    \newpage
    \subsection{CH\texorpdfstring{$\mathsf{_{3}}$}\ OH and H\texorpdfstring{$\mathsf{_{2}}$}\ O}
    $\left.\right.$
    \vfill
    \begin{figure}[h]
            \centering
            \begin{subfigure}{\textwidth}
                \centering
                \includegraphics[width=\textwidth]{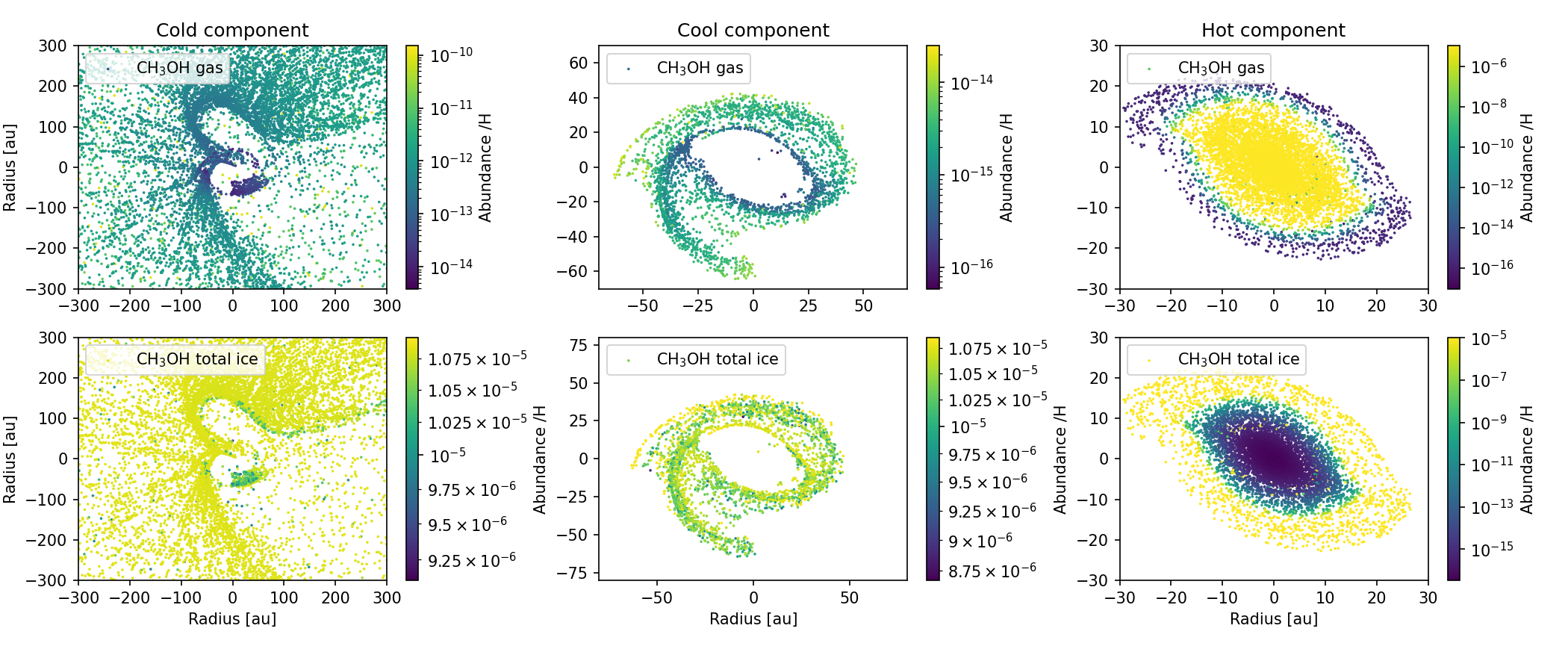}
                \caption{Spatial distribution of the tracer particles at the end of the simulation and their CH$_{3}$OH gas-phase (top row) and total ice-phase (bottom row) abundances  for the three temperature components: cold (left), cool (middle), and hot (right).}
            \end{subfigure}
            $\left.\right.$\\
            $\left.\right.$\\
            $\left.\right.$\\
            $\left.\right.$\\
            $\left.\right.$\\
            $\left.\right.$\\
            $\left.\right.$
            \begin{subfigure}{\textwidth}
                \centering
                \includegraphics[width=\textwidth]{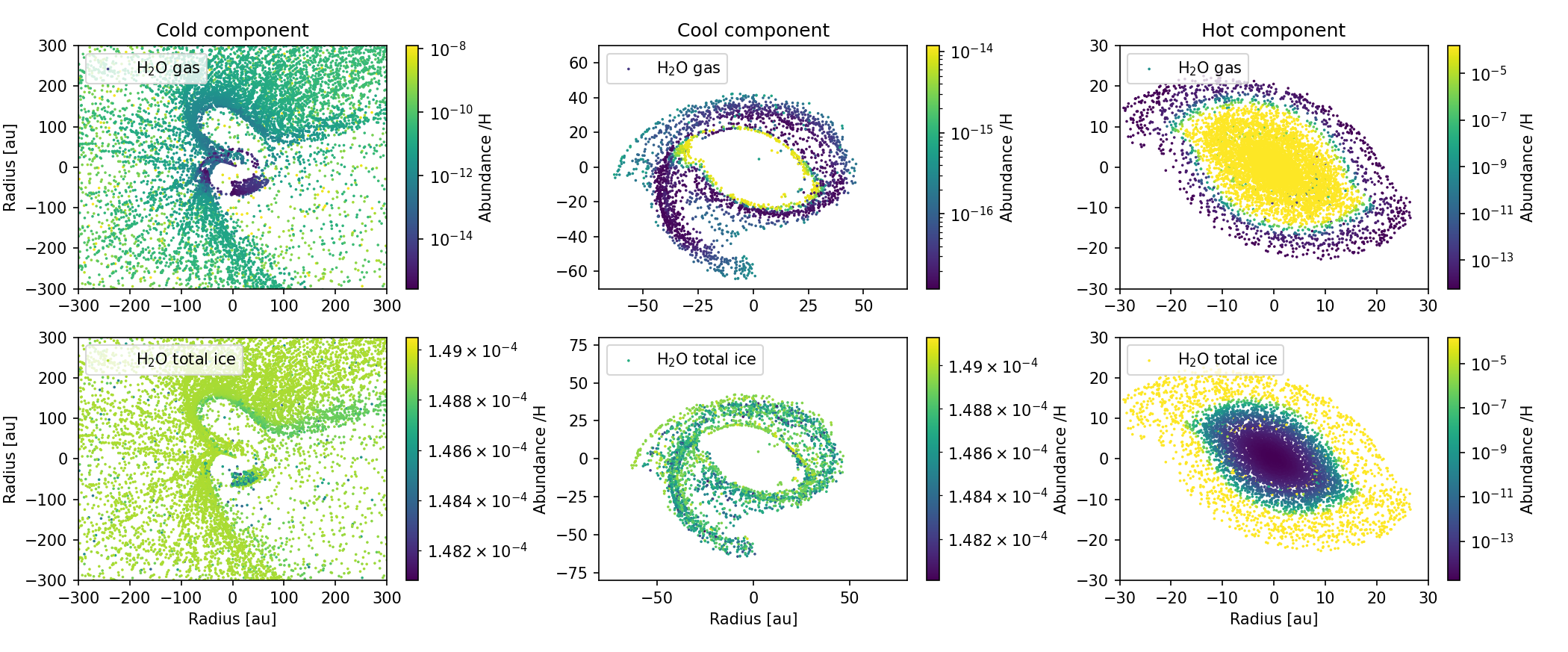}
                \caption{Spatial distribution of the tracer particles at the end of the simulation and their H$_{2}$O gas-phase (top row) and total ice-phase (bottom row) abundances  for the three temperature components: cold (left), cool (middle), and hot (right).}
            \end{subfigure}
        \end{figure}
        \vfill
    \newpage
    \subsection{CS and H\texorpdfstring{$\mathsf{_{2}}$}\ S}
    $\left.\right.$
    \vfill
    \begin{figure}[h]
            \centering
            \begin{subfigure}{\textwidth}
                \centering
                \includegraphics[width=\textwidth]{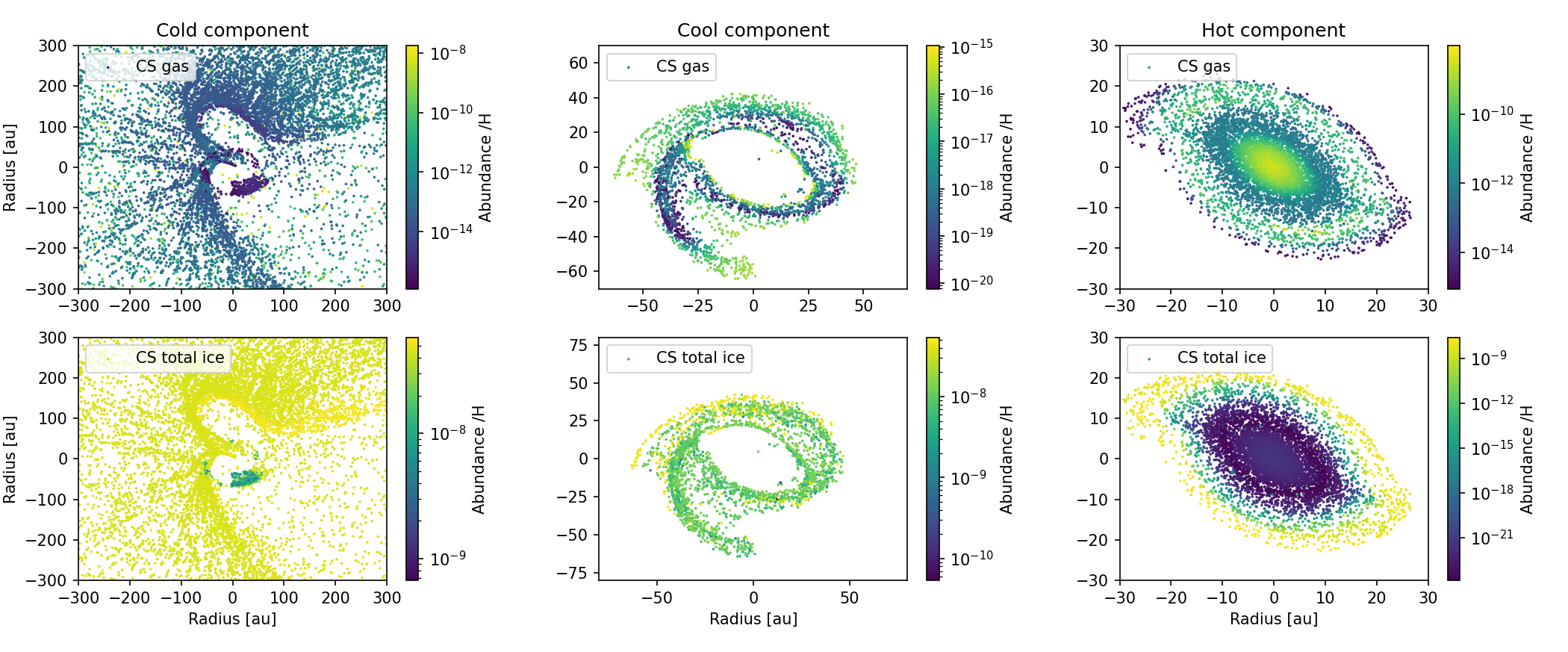}
                \caption{Spatial distribution of the tracer particles at the end of the simulation and their CH$_{3}$OH gas-phase (top row) and total ice-phase (bottom row) abundances  for the three temperature components: cold (left), cool (middle), and hot (right).}
            \end{subfigure}
            $\left.\right.$\\
            $\left.\right.$\\
            $\left.\right.$\\
            $\left.\right.$\\
            $\left.\right.$\\
            $\left.\right.$\\
            $\left.\right.$
            \begin{subfigure}{\textwidth}
                \centering
                \includegraphics[width=\textwidth]{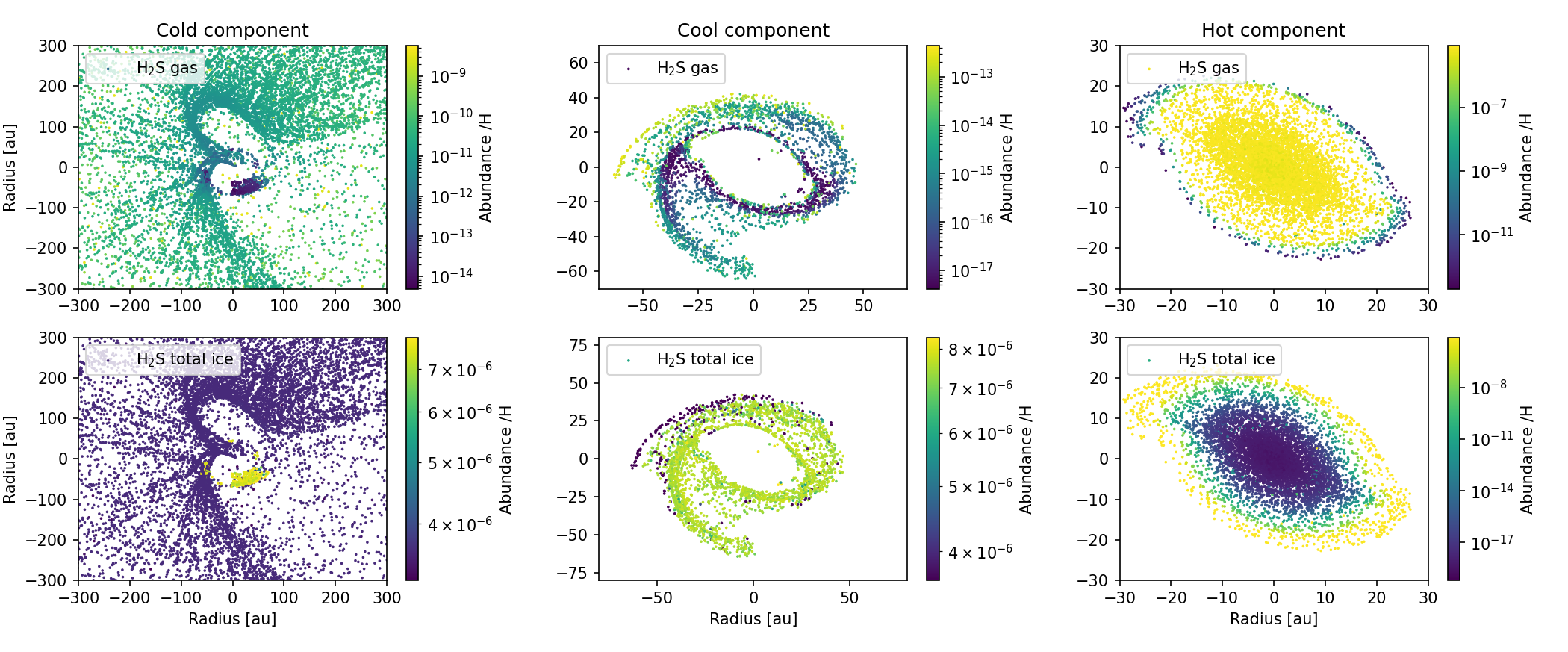}
                \caption{Spatial distribution of the tracer particles at the end of the simulation and their H$_{2}$O gas-phase (top row) and total ice-phase (bottom row) abundances  for the three temperature components: cold (left), cool (middle), and hot (right).}
            \end{subfigure}
        \end{figure}
        \vfill
        \newpage
    \subsection{SO and SO\texorpdfstring{$\mathsf{_{2}}$}\ }
    $\left.\right.$
    \vfill
    \begin{figure}[h]
            \centering
            \begin{subfigure}{\textwidth}
                \centering
                \includegraphics[width=\textwidth]{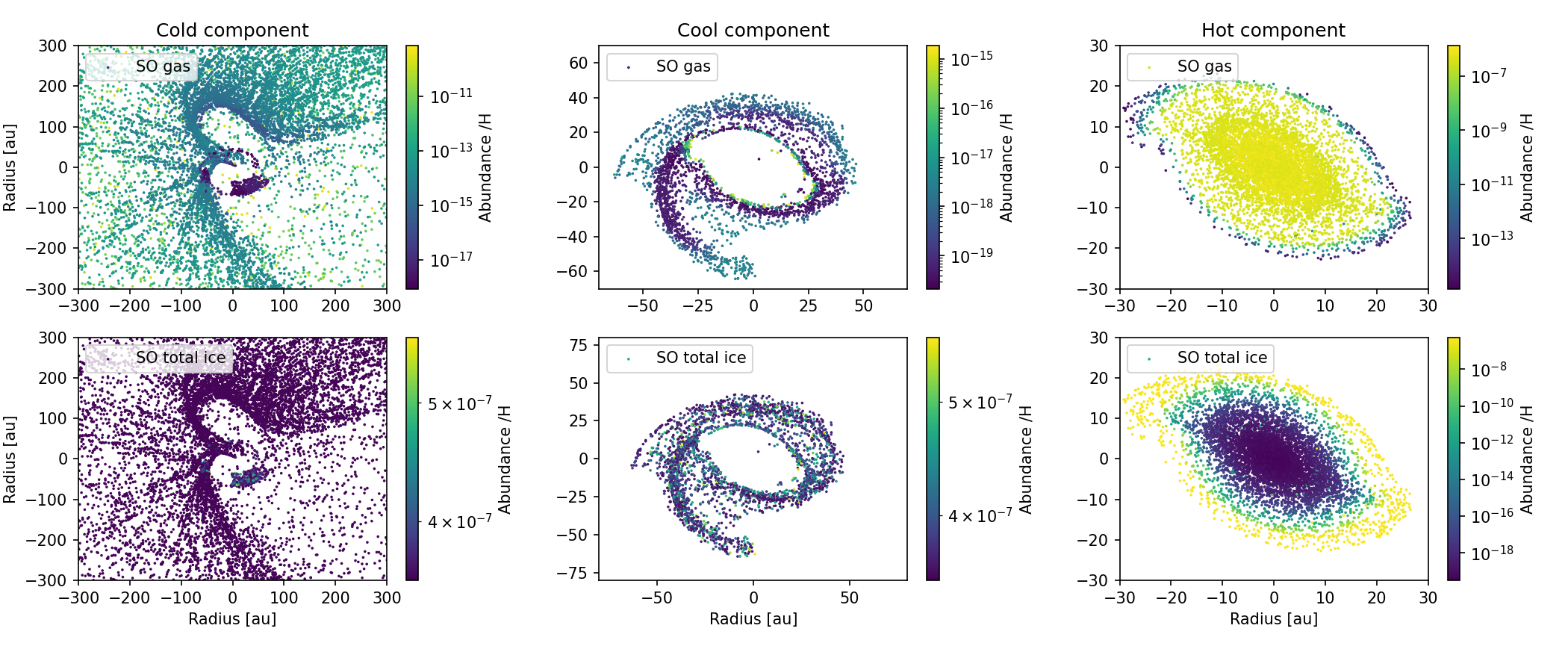}
                \caption{Spatial distribution of the tracer particles at the end of the simulation and their SO gas-phase (top row) and total ice-phase (bottom row) abundances  for the three temperature components: cold (left), cool (middle), and hot (right).}
            \end{subfigure}
            $\left.\right.$\\
            $\left.\right.$\\
            $\left.\right.$\\
            $\left.\right.$\\
            $\left.\right.$\\
            $\left.\right.$\\
            $\left.\right.$
            \begin{subfigure}{\textwidth}
                \centering
                \includegraphics[width=\textwidth]{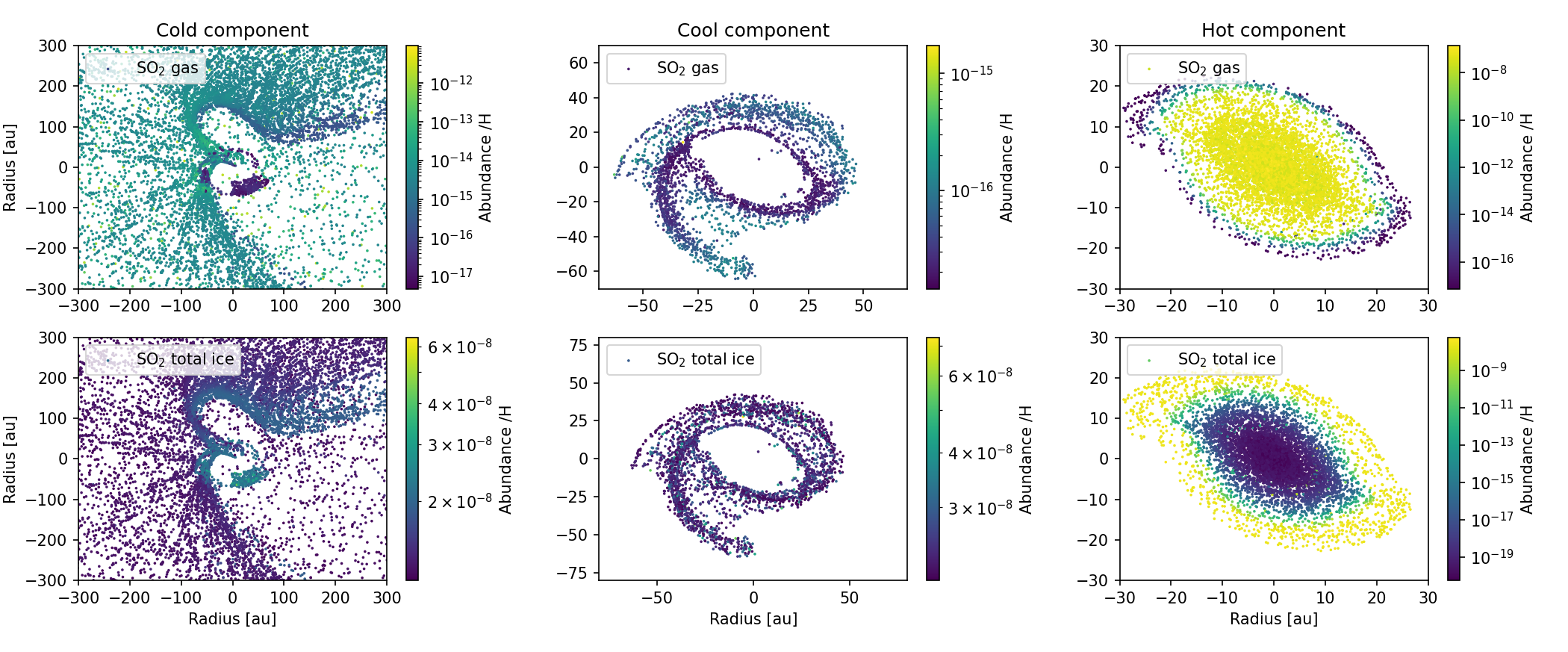}
                \caption{Spatial distribution of the tracer particles at the end of the simulation and their SO$_{2}$ gas-phase (top row) and total ice-phase (bottom row) abundances  for the three temperature components: cold (left), cool (middle), and hot (right).}
            \end{subfigure}
        \end{figure}
        \vfill
        \newpage
    \subsection{OCS and HCN}
    $\left.\right.$
    \vfill
    \begin{figure}[h]
            \centering
            \begin{subfigure}{\textwidth}
                \centering
                \includegraphics[width=\textwidth]{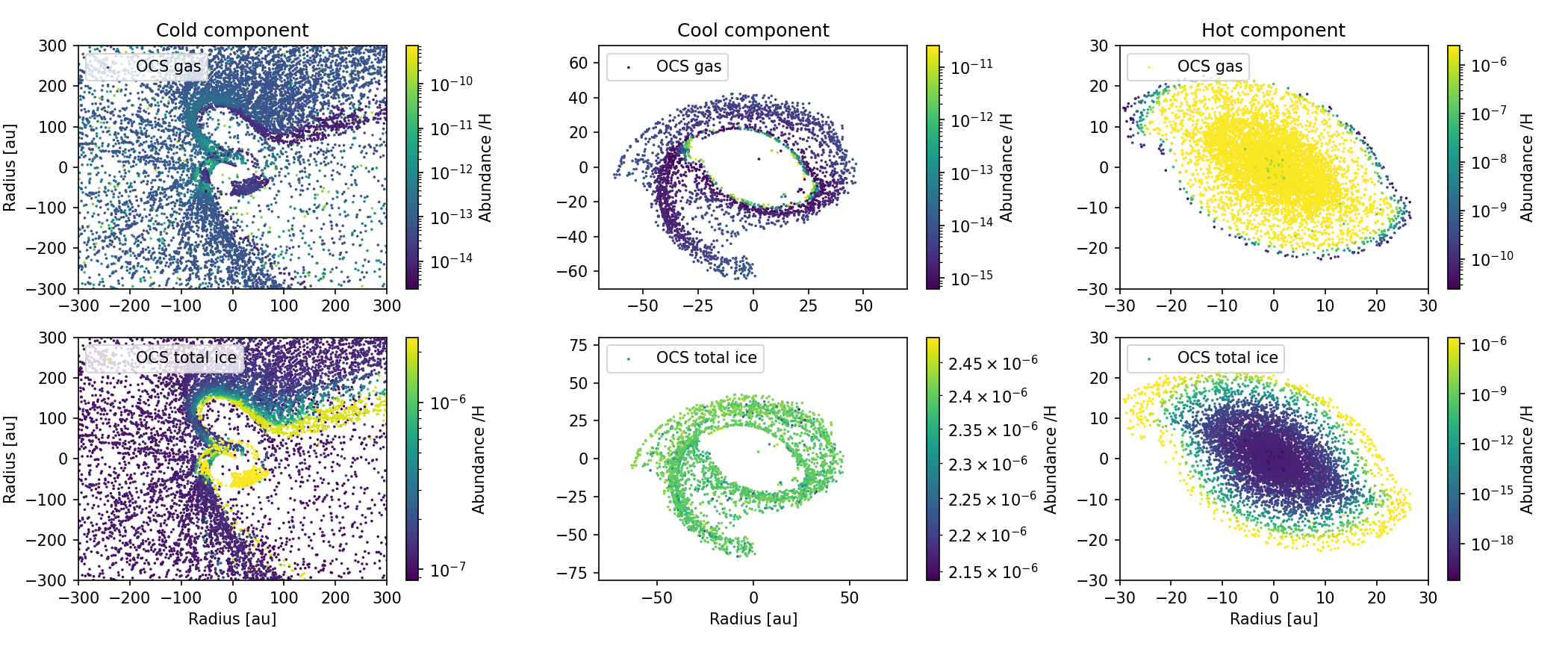}
                \caption{Spatial distribution of the tracer particles at the end of the simulation and their OCS gas-phase (top row) and total ice-phase (bottom row) abundances  for the three temperature components: cold (left), cool (middle), and hot (right).}
            \end{subfigure}
            $\left.\right.$\\
            $\left.\right.$\\
            $\left.\right.$\\
            $\left.\right.$\\
            $\left.\right.$\\
            $\left.\right.$\\
            $\left.\right.$
            \begin{subfigure}{\textwidth}
                \centering
                \includegraphics[width=\textwidth]{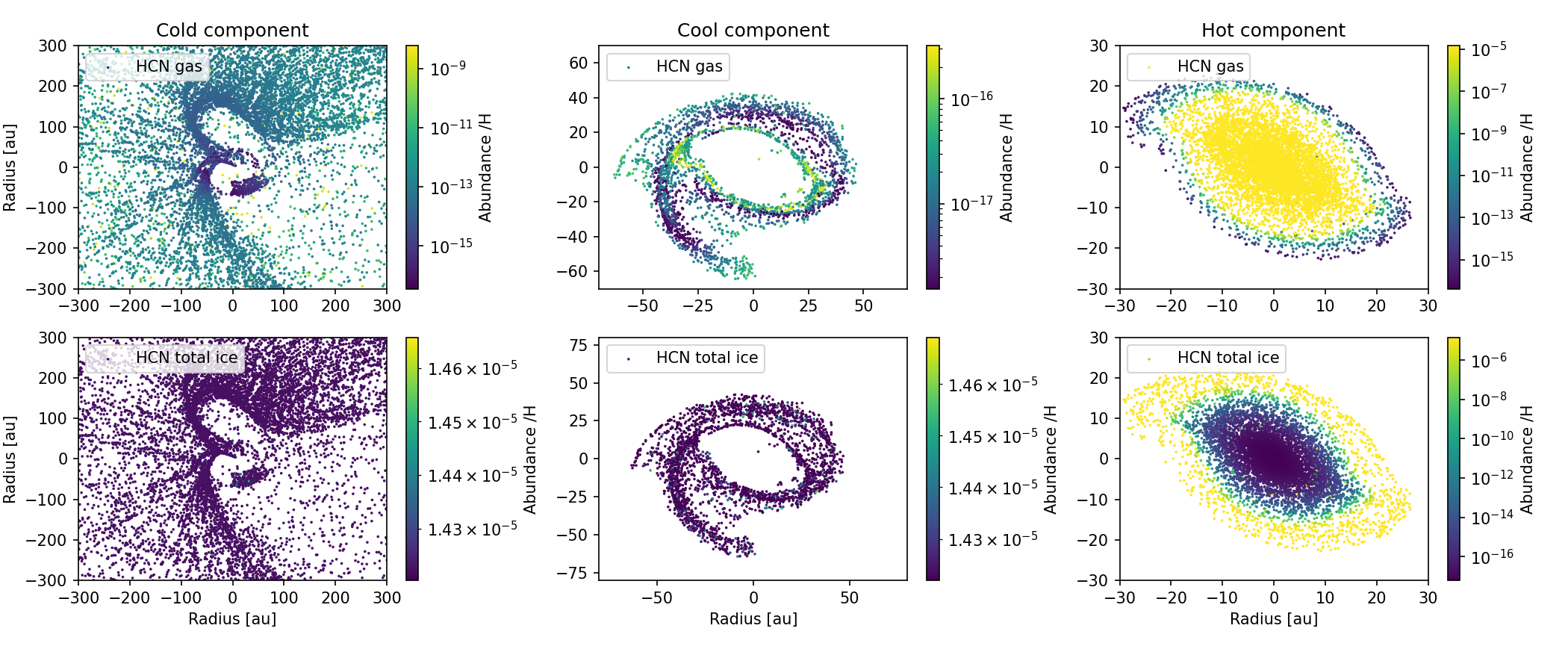}
                \caption{Spatial distribution of the tracer particles at the end of the simulation and their HCN gas-phase (top row) and total ice-phase (bottom row) abundances  for the three temperature components: cold (left), cool (middle), and hot (right).}
            \end{subfigure}
        \end{figure}
        \vfill
         \newpage
    \subsection{HNC, HCO\texorpdfstring{$\mathsf{^{+}}$}\ , and N\texorpdfstring{$\mathsf{_{2}}$}\ H\texorpdfstring{$\mathsf{^{+}}$}\ }
    $\left.\right.$
    \vfill
    \begin{figure}[h]
            \centering
            \begin{subfigure}{\textwidth}
                \centering
                \includegraphics[width=\textwidth]{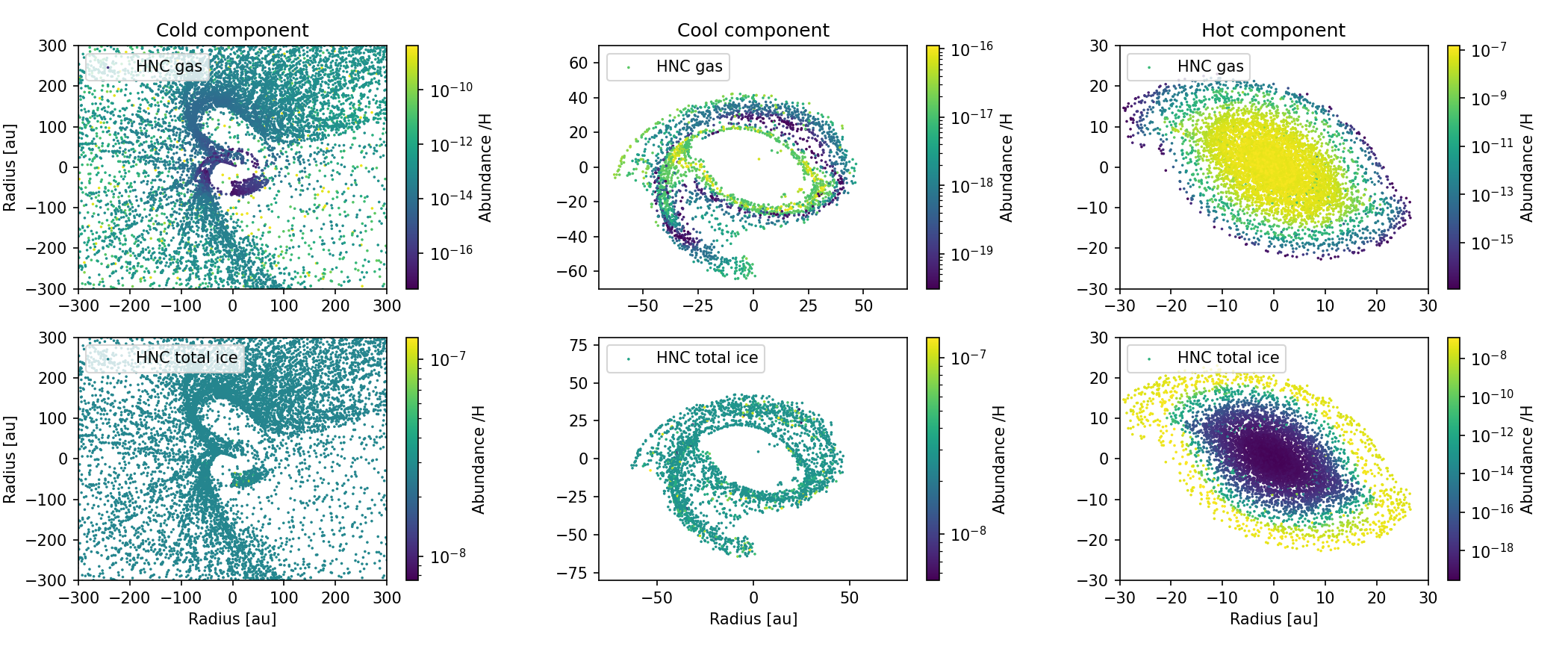}
                \caption{Spatial distribution of the tracer particles at the end of the simulation and their HNC gas-phase (top row) and total ice-phase (bottom row) abundances  for the three temperature components: cold (left), cool (middle), and hot (right).}
            \end{subfigure}
            $\left.\right.$\\
            $\left.\right.$\\
            $\left.\right.$\\
            $\left.\right.$\\
            $\left.\right.$\\
            $\left.\right.$\\
            $\left.\right.$
            \begin{subfigure}{\textwidth}
                \centering
                \includegraphics[width=\textwidth]{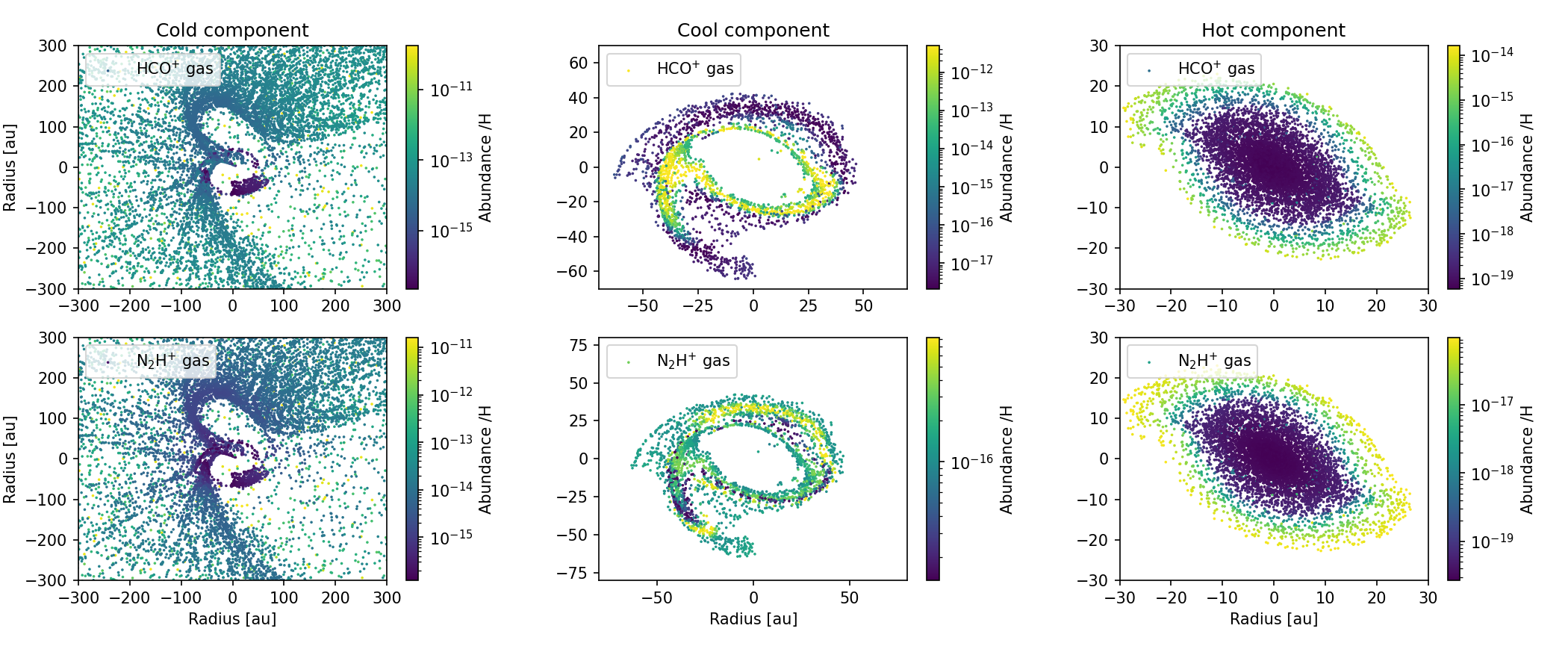}
                \caption{Spatial distribution of the tracer particles at the end of the simulation and their HCO$^{+}$ (top row) and N$_{2}$H$^{+}$ (bottom row) gas-phase abundances for the three temperature components: cold (left), cool (middle), and hot (right).}
            \end{subfigure}
        \end{figure}
        \vfill
    \newpage
\end{appendix}
\end{document}